\documentclass[twocolumn]{aastex63}
\usepackage{graphicx,float} 
\usepackage{booktabs}
\usepackage{longtable}
\usepackage{float}
\usepackage[utf8]{inputenc}
\usepackage{ae}
\usepackage{aecompl}
\usepackage{amsmath, amssymb, amsfonts, braket, gensymb} 
\usepackage{newtxtext,newtxmath} 
\usepackage{enumitem}
\defcitealias{2013MNRAS.429.1949A}{AMR13}
\DeclareMathOperator{\sech}{sech}
\renewcommand{\textcolor}[2]{#2}

\shorttitle{Buckling Bars}
\shortauthors{Xiang et al.}

\begin{document}
	
\title{Buckling bars in \textcolor{red}{nearly} face-on galaxies observed with MaNGA}
\correspondingauthor{Katherine M. Xiang}
\email{kxiang@g.harvard.edu, katherine.m.xiang@gmail.com}

\correspondingauthor{David M. Nataf}
\email{dnataf1@jhu.edu, david.nataf@gmail.com}

\author{Katherine M. Xiang}
\affil{Center for Astrophysical Sciences and Department of Physics and Astronomy,
The Johns Hopkins University, 
Baltimore, MD 21218, USA}

\affil{Department of Physics, Harvard University, Cambridge, MA 02138, USA.}

\author{David M. Nataf}
\affiliation{Center for Astrophysical Sciences and Department of Physics and Astronomy,
The Johns Hopkins University,
Baltimore, MD 21218, USA}

\author{E. Athanassoula}
\affiliation{Aix Marseille Universit\'e, CNRS,  CNES, LAM, Marseille, France}

\author{Nadia L. Zakamska}
\affiliation{Center for Astrophysical Sciences and Department of Physics and Astronomy,
The Johns Hopkins University,
Baltimore, MD 21218, USA}

\author{Kate Rowlands}
\affiliation{Space Telescope Science Institute, 
3700 San Martin Drive, 
Baltimore, MD 21218, USA}

\author{Karen Masters}
\affiliation{Department of Physics and Astronomy,
Haverford College,
370 Lancaster Ave, Haverford, PA 19041, USA}

\author{Amelia Fraser-McKelvie}
\affiliation{School of Physics and Astronomy,
University of Nottingham,
University Park, Nottingham, NG7 2RD, UK}

\author{Niv Drory}
\affiliation{McDonald Observatory,
University of Texas at Austin,
1 University Station, Austin, TX 78712, USA}

\author{\textcolor{red}{Katarina Kraljic}}
\affiliation{Aix Marseille Universit\'e, CNRS,  CNES, LAM, Marseille, France}

	\begin{abstract}
		Over half of disk galaxies are barred, yet the mechanisms for bar formation and the life-time of bar buckling remain poorly understood. In simulations, a thin bar undergoes a rapid (<1 Gyr) event called ``buckling,'' during which the inner part of the bar is asymmetrically bent out of the galaxy plane and eventually thickens, developing a peanut/X-shaped profile when viewed side-on. Through analyzing stellar kinematics of N-body model snapshots of a galaxy before, during, and after the buckling phase, we confirm a distinct quadrupolar pattern of out-of-plane stellar velocities in near\textcolor{red}{ly} face-on galaxies. This kinematic signature of buckling allows us to identify five candidates of currently buckling bars \textcolor{red}{among 434 barred galaxies in the Mapping Nearby Galaxies at Apache Point Observatory (MaNGA) Survey}, an integral field unit (IFU) spectroscopic survey that measures the composition and kinematic structure of nearby galaxies. The frequency of buckling events detected is consistent with the 0.5-1 Gyr timescale predicted by simulations. The five candidates we present more than \textcolor{red}{double} the total number of candidate buckling bars, and are the only ones found using the kinematic signature.
		\newline\vspace{2.0em}
	\end{abstract}
  

	\section{Introduction} \label{sec:Introduction}
	
Galactic bars are elongated, smooth stellar systems in the inner parts of disk galaxies\textcolor{red}{. They} can be morphologically approximated as triaxial ellipsoids, \textcolor{red}{and} sometimes have a super-imposed peanut/X-shaped structure when viewed side-on \citep{1981A&A....96..164C}. Within the local universe, it is estimated that roughly two thirds of disk galaxies have stellar bars \citep{1997ApJ...482L.135M,2000AJ....119..536E,2007ApJ...657..790M,2008ApJ...675.1194B,2008ApJ...675.1141S,2009A&A...495..491A,2019ApJ...872...97L}, though the precise number is uncertain due to differences in factors such as the classification scheme and the details of the methodology. For example, \citet{1997ApJ...482L.135M} found that 55\% of the galaxies classified as ``un-barred" in the \textit{Revised Shapley-Ames Catalog of Bright Galaxies} \citep{1981rsac.book.....S} were barred in 2.1${\mu}$m data. 
	
The presence and strength of bars are major variables of Hubble's galaxy sequence \citep{1936rene.book.....H}, still the most commonly used classification scheme for galaxies. It is thus not surprising that an understanding of galaxy evolution requires an understanding of bars, \textcolor{red}{and that} the relationships between bars and other galaxy properties have been subject to extensive research. \textcolor{red}{ Observationally, the probability} of a galaxy hosting a bar is a decreasing function of its specific star formation rate \citep{2011MNRAS.411.2026M, 2013ApJ...779..162C}. The bar fraction is minimized for spirals with $\log({M_{\rm{stellar}}/M_{\odot}}) \approx 10.20$, and then increases for both lower and higher mass spiral galaxies \citep{2010ApJ...714L.260N}. High redshift galaxies are less likely to host strong bars: \citet{2008ApJ...675.1141S} find that the barred fraction in a sample of luminous spirals declines from 65\% at $z=0$ to 20\% at $z \approx 0.84$, and \citet{2014MNRAS.438.2882M} find that the bar fraction in a sample of visually selected disk galaxies decreases from 22\% at $z=0.4$ to 11\% at $z=1.0$.

N-body models have yielded numerous, detailed predictions for \textcolor{red}{evolution,} kinematics and photometry \textcolor{red}{of bars} in observed galaxies. \textcolor{red}{They demonstrate that bars grow and evolve through their interaction with the stellar and dark matter haloes of their host galaxies (\citealt{1985MNRAS.213..451W, deba98,deba00,2003LNP...626..313A,2003MNRAS.341.1179A}; \citealt{2013MNRAS.429.1949A}, hereafter AMR13) and with other galaxies \citep{1990A&A...230...37G,10.1093/mnras/stu1846}.} Thus far, the match between the predictions from simulations and the observations has been strikingly impressive, both for photometric features in field galaxies \textcolor{red}{\citep{10.1093/mnras/stv2231,2017A&A...598A..10L,2017ApJ...835..252S}}, and in the Milky Way \citep{2015MNRAS.447.1535N,2017MNRAS.467L..46A,2017MNRAS.465.1621P,2019MNRAS.490.4740B}. \textcolor{red}{Comparisons between observations of the Milky Way bar and N-body simulations can now be used in Galactic archaeology and for probing the Galactic merger history and bulge formation \citep{shen10, ness13, 2017MNRAS.468.2058E, dima19}. \citet{2005ApJ...626..159B} analyzed N-body simulations of galactic
discs from \citet{2003MNRAS.341.1179A}, viewed them edge-on, and found a number of characteristic kinematic bar features. One such interesting feature is that the mean and skewness of the radial profiles from the line-of-sight velocity distributions are positively correlated over the projected bar length. This was also found by \cite{Chung_2004} for a sample of 30 field galaxies from the work of \cite{1999AJ....118..126B}, as well as for the metal rich stars in the Milky Way \citep{2016ApJ...832..132Z}. Such diagnostics are reviewed by \citet{2013seg..book..305A,2016ASSL..418..391A}.}

In this paper, we study a prediction from N-body models of a specific phase of bar evolution, that of the \textit{buckling} of the bar \citep{1981A&A....96..164C, 1990A&A...233...82C,1991Natur.352..411R}. \textcolor{red}{Buckling is one of several dynamical processes that have been discussed in the literature as a means for an in-plane bar to evolve into a boxy/peanut bulge \citep{2014MNRAS.437.1284Q,2014ASPC..480..157P,2020MNRAS.495.3175S}.} The buckling phase is due to a violent dynamical instability, during which an initially thin bar bends asymmetrically out of its host galaxy's plane over a short period of time and then settles to a vertically thicker \textcolor{red}{symmetric} configuration. The bar settles to having a boxy/peanut shape when viewed edge-on. \textcolor{red}{The changes in the bar's structure can be large: the first buckling event of the galaxy simulation} studied by \citet{2006ApJ...637..214M} culminates in the height of the bar nearly doubling, and the size of the bar's semimajor axis shrinking by nearly a half. 

Despite the \textcolor{red}{near-}ubiquity of \textcolor{red}{buckling in simulations, few predictions for the buckling phase have been tested with direct observations. If buckling is short-lived, then finding galaxies in the buckling phase may be difficult. Some models predict longer-lived ($\sim$2-3 Gyr) secondary buckling events \citep{2006ApJ...637..214M, 2019A&A...629A..52L}, though the frequency of these events is lower \citep{2019MNRAS.485.1900S} than that of the first buckling. Note that this frequency depends on the strength of the bar \citep{2008IAUS..245...93A}.}  Only \textcolor{red}{three candidates have been identified so far: \citealt{2016ApJ...825L..30E}, who built on the work of \citealt{2013MNRAS.431.3060E}}, showed that the surface brightness profiles in the nearby face-on spirals NGC 3227 and NGC 4569 are consistent with predictions for buckling from N-body simulations. \textcolor{red}{One additional candidate has been identified by the same method and is described in \citet{Li_2017}.} Specifically, the central surface brightness isophotes of these bars show 180$^{\circ}$ symmetry with respect to reflection around the bar's minor axis, but not the major axis. \textcolor{red}{Another potentially testable prediction of the buckling simulations is that the buckling instability is expected to be inhibited in galaxies with a large gas fraction \citep{2006ApJ...645..209D, 2007ApJ...666..189B, 2009A&A...494...11W, 2010ApJ...719.1470V, 2019MNRAS.485.1900S}. Testing this and other predictions and evaluating the timescale for buckling requires a large sample of buckling candidates.} 

Here, we develop a different set of \textcolor{red}{buckling metrics}, focused on kinematics rather than surface brightness, by inspection of a suite of N-body models observed at different snapshots. We then \textcolor{red}{apply these metrics to} a sample of nearby galaxies observed by the Mapping Nearby Galaxies at Apache Point Observatory (MaNGA) survey \citep{2015ApJ...798....7B} and present five candidate galaxies with currently buckling bars.

The structure of this paper is as follows. In Section \ref{sec:Models}, we introduce the models that are used to develop our predictions, which we then proceed to quantify. In Section \ref{sec:Analysis}, we describe the sample of galaxies from the Mapping Nearby Galaxies at Apache Point Observatory (MaNGA) survey on which we test these predictions. \textcolor{red}{We then proceed to identify and evaluate buckling candidates from this sample}. We present a discussion of our findings in Section \ref{sec:Discussion}, and conclude in Section \ref{sec:Conclusion}. In this study, we use $H_0 = 100 \: (\textrm{km/s})/\textrm{Mpc}$.

    \section{Models} \label{sec:Models}
    
    \subsection{N-body Model}
    
The three simulation snapshots studied here are part of a larger suite of simulations run by \citetalias{2013MNRAS.429.1949A}. \textcolor{red}{These simulations include a live halo, \textit{i.e.} a halo where the dark matter particles can exchange energy and angular momentum with the baryons, a realistic prescription for gas physics, and a suite of initial conditions with varying gas fractions and halo shape.} We chose the simulation producing the strongest bar and peanut, in order to have the clearest example of the buckling signature with which to compare to observations; thus, the initial conditions had a spherical halo and no gas. Bar strength was determined by considering the $m=2$ Fourier component of the face-on mass distribution \citep{2003MNRAS.341.1179A}. \textcolor{red}{The simulation used in this work is referred to as run 101 and GTR-101 in  \citetalias{2013MNRAS.429.1949A} and \citet{2015MNRAS.450.2514I}, respectively, and these two studies describe the physics, initial conditions, and numerical methods in greater detail.}

Briefly, the \textcolor{red}{initial conditions of the simulations were constructed using the iterative method \citep{2009MNRAS.392..904R}, and were run} using a version of GADGET-3  \citep*{2001NewA....6...79S,2002MNRAS.333..649S,2005MNRAS.364.1105S}. The dark matter and the stars are followed by N-body particles, and gravity is calculated with a TREE code. The models have a 50 pc linear resolution, and a $2.5 \times 10^4 M_{\odot}$ mass resolution for the stars. There are 200,000 particles representing the stars. Two thousand output files are saved for this run, corresponding to different snapshots in time.

The stars are initially distributed as a disk, whose azimuthally-distributed average spatial distribution is given by: 
\begin{equation}
\rho_d (R, z) = \frac {M_d}{4 \pi h^2 z_0}~~\exp (- R/h)~~\sech^2 \Big(\frac{z}{z_0}\Big),
\end{equation}
\noindent
where $R$ is the cylindrical radius, $z$ is the vertical coordinate, $M_d = 5 \times 10^{10} M_{\odot}$ is the disk mass, $h =3$~kpc is the disk radial scale length, and $z_0=0.6$~kpc is the disk vertical scale thickness. \textcolor{red}{The initial radial velocity dispersion of the particles is initialized to:}
\begin{equation}
\sigma_{R}(R) = 100 \exp (-R/3h)\, \rm{km}\,\rm{s}^{-1}
\end{equation}

\textcolor{red}{The halo has been built so as to have an initial distribution function that is spherical, with a radial density profile distributed as}:
\begin{equation}
\rho_h (R) = \frac {M_h}{2 \pi^{3/2}}~\frac{\alpha}{r_{c}}~\frac{\exp (-r^{2}/r_{c}^2)}{r^2+\gamma^2},
\end{equation}
\noindent
\textcolor{red}{where $r$ is the spherical radius, $M_{h}=2.5 \times 10^{11} M_{\odot}$ is the mass of the halo, and $\gamma=1.5$~kpc and $r_{c}=$~30 kpc are
the halo core and cut-off radii, respectively. The parameter $\alpha$ is a normalization
constant. }

\subsection{Snapshots from N-body Simulation}\label{sec:Methods}

We select three snapshots for our present investigation, at 2.00 Gyr, 3.53 Gyr, and 4.50 Gyr. We choose these particular snapshots so as to study the bar of the galaxy in its pre-buckling state, during the buckling, and in its post-buckling state, \textcolor{red}{respectively.} \textcolor{red}{Figures 4 and 5 of \citetalias{2013MNRAS.429.1949A} show further snapshots at 6 and 10 Gyr, i.e. well after the buckling. Snapshots from the earlier times, more relevant to this work, will be given in our analysis in the following sections. Figures 7 and 8 of \citetalias{2013MNRAS.429.1949A} also show the time evolution of the bar strength $A_2$. Here the buckling epoch can be easily identified from the sharp drop in the bar strength parameter in the top-left panels of both these figures. The time evolution of the bar strength is also plotted in Figure 1 of \citet{2015MNRAS.450.2514I}. These figures demonstrate that buckling corresponds to a short-lived sharp decrease of the bar strength. This starts somewhat after 3 Gyr, and by
4.8 Gyr, it has recovered to its peak pre-buckling value.}

\begin{figure}[h]
    \includegraphics[width=1.0\linewidth]{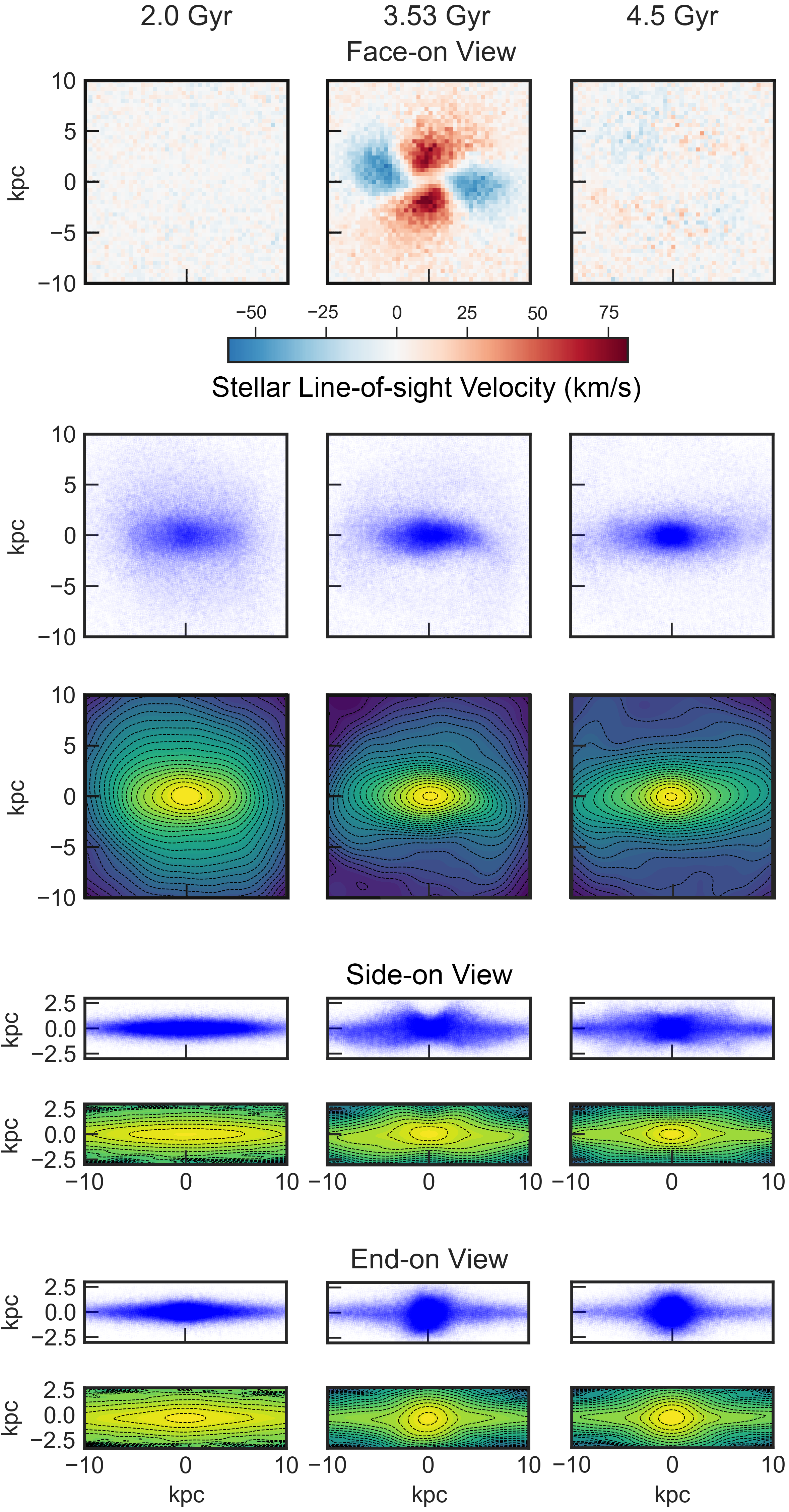}
    \caption{Top row displays plots of mean stellar line of sight velocity of the simulated galaxy viewed face-on, with 2500 total bins in a 400 kpc$^2$ map. The particles of the N-body model with the same model view are shown in the second row, and the surface density and isophotes are plotted in the third row. The side-on maps of model particles and surface densities are shown in the next two rows, and the end-on\textsuperscript{\textcolor{red}{a}} maps are shown in the final two rows. Note that from the \textcolor{red}{side-on and edge-on views, the buckling bar is asymmetric with respect to the z=0 plane, while viewed end-on it is asymmetric with respect to the plane perpendicular to the bar minor axis. 25 logarithmically spaced isodensities are plotted for all density maps.} \vspace{1em}}
    \label{figure:buckling_faceon}
    \footnotesize{\textcolor{red}{\textsuperscript{a}We call \textit{side-on} the edge-on view in which the line of sight is along the bar minor axis, and \textit{end-on} the edge-on view in which the line of sight is along the bar major axis.}}
    \end{figure}

We investigate the mean line-of-sight velocity of disk particles \textcolor{red}{in each snapshot.} We bin the 200,000 particles using equal spatial intervals (0.4 by 0.4 kpc square bins in Figure \ref{figure:buckling_faceon}), and we plot the average line-of-sight velocity of particles in each bin. Maps are generated for all three snapshots from a variety of viewing and inclination angles in an effort to observe a signal of bar buckling \textcolor{red}{based on the} simulated stellar velocities. 
    
In Figure \ref{figure:buckling_faceon}, we show the face-on maps of the mean stellar line-of-sight velocities, as well as the particle distributions and stellar density profiles for three different orientations, for each of the three snapshots. For a smoothly rotating disk viewed face-on, one would expect little to no \textcolor{red}{line-of-sight} component of velocity, as seen in the pre-buckling and post-buckling snapshots. However, at the time of buckling, we observe a significant out-of-plane component velocity signature, resembling a quadrupole in stellar velocities with a diameter of approximately 18 kpc along the bar major axis, as seen in the middle panel of the top row of Figure \ref{figure:buckling_faceon}. The velocity signal is most distinct \textcolor{red}{when viewed} directly face-on, which we define to be $i=0^{\rm o}$, i.e., \textcolor{red}{when viewing} the galaxy with a line-of-sight that is parallel to the galaxy's angular momentum vector. (We elaborate on the coordinate system in Section \ref{sec:coords}.) For this model, the peak velocity amplitude of the buckling signal is $\sim$75 km/s (measured from a face-on view), which is approximately one-third of that of the galaxy's rotation signal (measured from an edge-on view).
    
This kinematic signature has also been found by \citet{2019A&A...629A..52L} \textcolor{red}{through analysis of an isolated disk galaxy model}; the middle panel of Figure 6 depicts a similar quadrupolar pattern in stellar velocities. The same signal has also been found and discussed in \citet{fragkoudi2019chemodynamics}, which uses magneto-hydrodynamical Milky Way-mass galaxy simulations (the Auriga suite, \citealt{2018MNRAS.474.3629G,2019MNRAS.490.4786G}). The quadrupolar signal is displayed and discussed in their Figures 10 and B1. Thus, the quadrupolar kinematic signature of buckling is a robust \textcolor{red}{and quantifiable} feature of the buckling phase, and is independent of simulation specifics. \textcolor{red}{We have also examined the line-of-sight velocity dispersion maps, but unlike in velocities, there was no unambiguously clear signal associated with buckling.} 
    
\begin{figure}
    \includegraphics[width=1.0\linewidth]{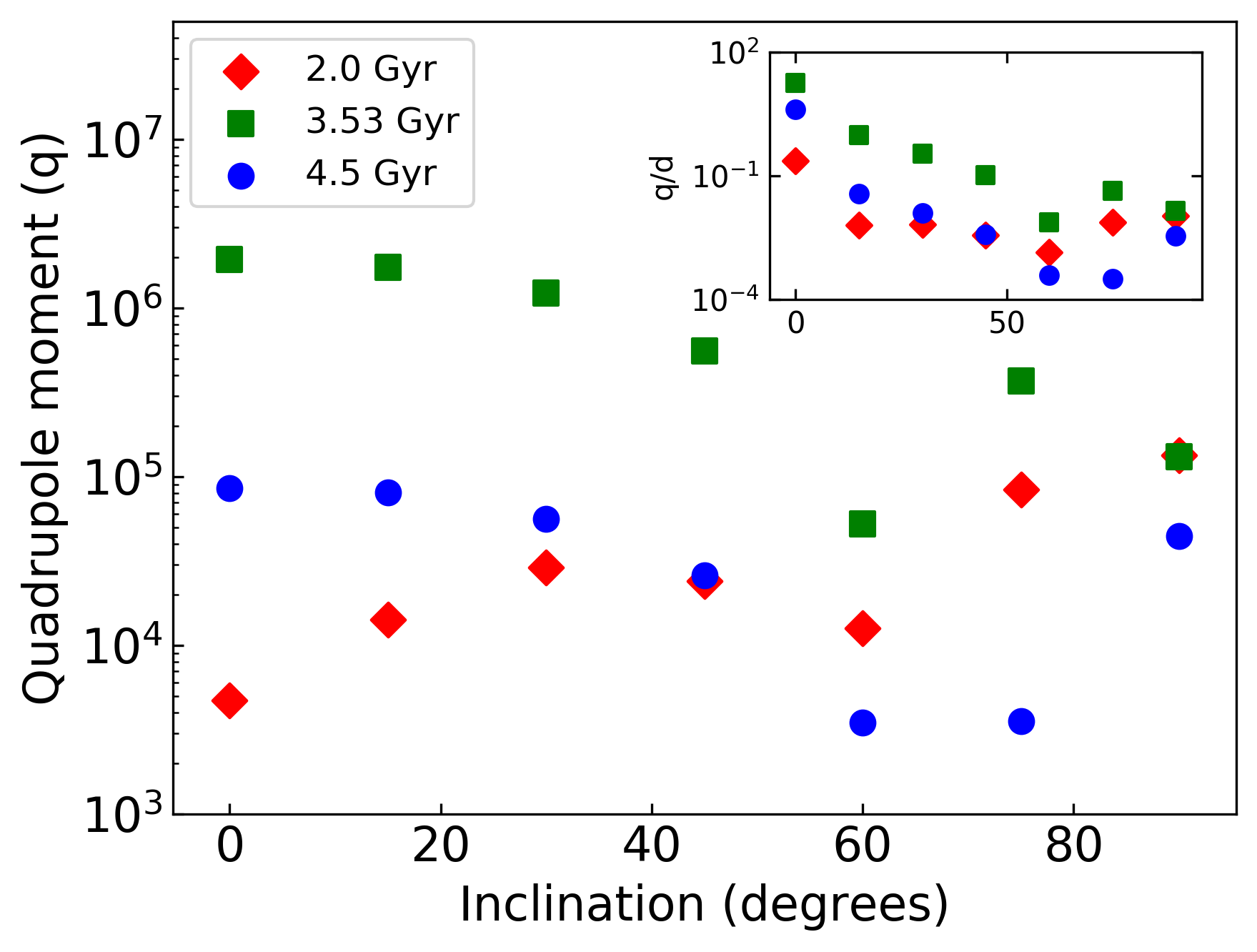}
    \caption{Quadrupole moment for the three simulation snapshots, as defined by Equation \ref{eq:quadmoment}. The quadrupole moment of the buckling snapshot \textcolor{red}{decreases with increasing inclination angle,} consistent with visual observation such as in the second column of Figure \ref{figure:orientations_horizontal}. \textcolor{red}{At 60 - 70 degrees, the quadrupole moment for the buckling snapshot still remains several times higher than either of the non-buckling snapshots. Inset: quadrupole moment normalized by dipole moment.}}
    \label{figure:quadmoment}
    \end{figure}

At the end of the buckling phase (Figure \ref{figure:buckling_faceon}, right column), a slight quadrupolar pattern indicates that vestiges of the buckling signal are still present nearly a full billion years after peak buckling. Therefore, we seek a metric to quantify the strength of buckling and isolate the buckling phase. To this end, we consider the quadrupole moment of the map, $q$:

    \begin{equation}
        q  = \sum_{n = 0}^{N} V_{\textrm{los, }n}\cos(2(\alpha_n-\alpha_0))\textcolor{red}{,}
    \label{eq:quadmoment}
    \end{equation}
    
    \noindent where we sum over all model particles up to the radius of the bar, and $V_\mathrm{los}$ is the line-of-sight velocity of each model particle. $\alpha_0$ and $\alpha$ are, respectively, the bar angle and the position angle of each model particle in the image plane, both measured counterclockwise from east. Values of $q$ are plotted for a variety of inclinations and for all three simulation snapshots in Figure \ref{figure:quadmoment}. \textcolor{red}{We also plot $q/d$, the quadrupole moment normalized by the dipole moment, which takes into account the signal strength compared to the in-plane stellar velocities. We obtain qualitatively similar results. In summary, our primary observable is the quadrupole moment of the velocity field. We note that in this simulation, the bar exhibits a strong in-plane bending as well (shown in the isodensities in Figure \ref{figure:buckling_faceon}), and it is not known whether this in-plane bending of the bar is typical of buckling. However, the quadrupolar signature is not unique to the bar that is present in this model. Estimates of $q$ from different simulation snapshots in the literature (e.g. \citealt{2019A&A...629A..52L, fragkoudi2019chemodynamics}) are similar to that in our simulation.}
    
    \subsection{Coordinate System}
    \label{sec:coords}
    For comparison to observational data, we determine the appearance of the buckling signal from various viewing angles. The position of the observer is specified by two angles: a polar angle and an azimuthal angle. The polar angle is specified by $\theta$, and ranges from $0^{\circ} - 180^{\circ}$. Both $\theta = 0^{\circ}$ and $\theta = 180^{\circ}$ correspond to an inclination of $i = 0^{\circ}$, for comparison for astronomical data. The azimuthal angle $\varphi$ also ranges from $0^{\circ}$ to $180^{\circ}$. A final angle $\gamma$ corresponds to rotations in the image plane. A face-on galaxy with a horizontal bar corresponds to ($\theta = 0^{\circ}$, $\varphi  = 0^{\circ}$, $\gamma = 0^{\circ}$), as in the top row of Figure \ref{figure:buckling_faceon}, or ($\theta = 180^{\circ}$, $\varphi  = 0^{\circ}$, $\gamma = 0^{\circ}$), which corresponds to the other side of the disk. 
    
    We transform the coordinates of particles in the N-body model and obtain average line-of-sight velocities for each bin. To determine the transformed projected bar angle, we define a unit vector of the bar (1,0,0) and transform it using the same rotation matrices. Then, we project this vector into the plane of the image for comparison of the projected bar angle to observational data. We take the magnitude to obtain the projected simulation bar length.
    
\begin{figure*}
        \centering
        \includegraphics[width= 1.0\linewidth]{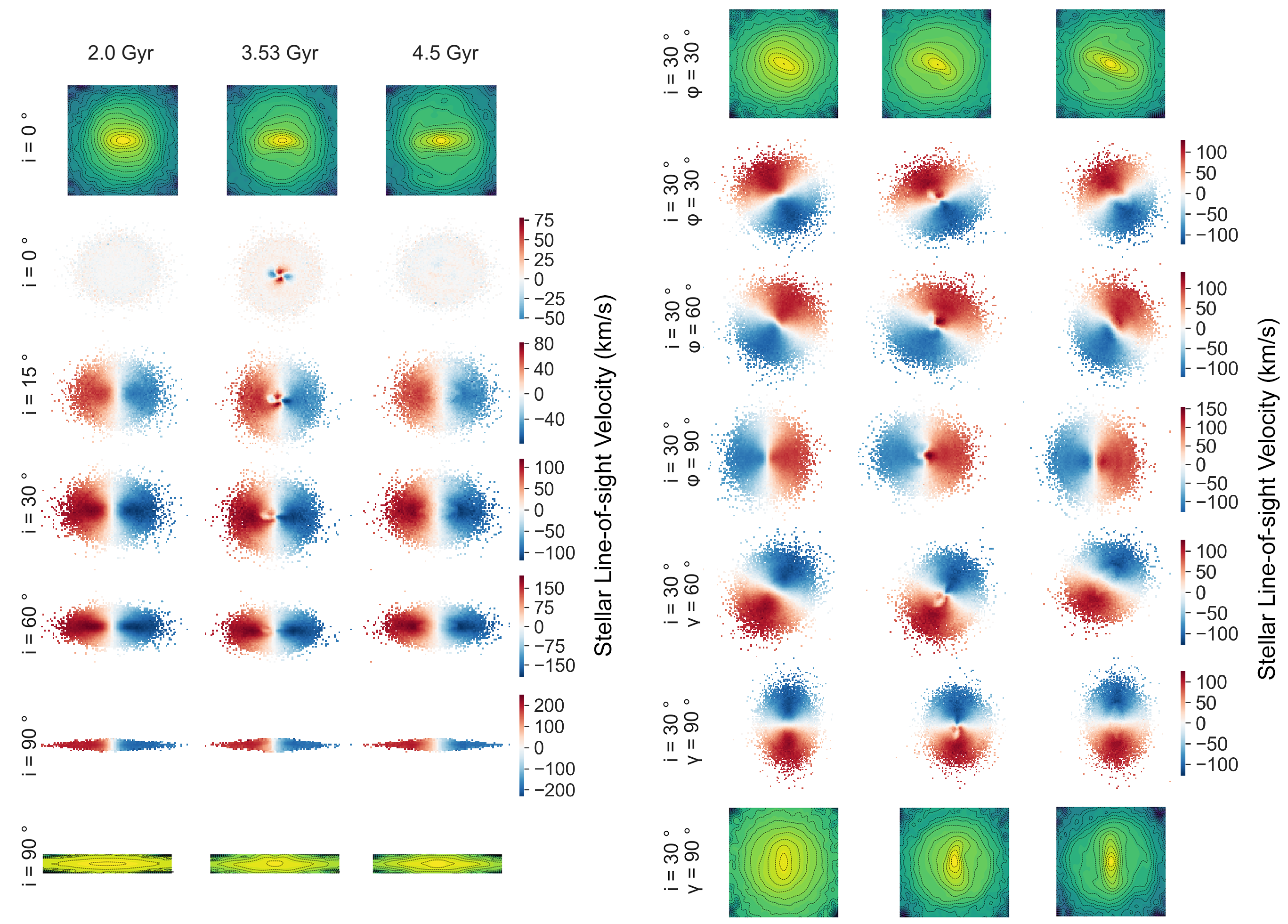}
        \caption{Simulation snapshots from different viewing angles, with square bins \textcolor{red}{of 0.4 kpc side length}. Left: Each row corresponds to the stellar velocity map of the model in a different orientation, and each column corresponds to a different time in galaxy evolution: 2.0, 3.53, and 4.5 Gyr\textcolor{red}{, respectively}. Top and bottom rows in each column are surface density maps with logarithmic\textcolor{red}{ally spaced contours}, to guide the eye. The rows correspond to $i = 0^{\circ}$, $15^{\circ}$, $30^{\circ}$, $60^{\circ}$, $90^{\circ}$ with $\varphi = 0^{\circ}$ and $\gamma = 0^{\circ}$. The view proceeds from a face-on view to an edge-on one, with the bar major axis perpendicular to the line-of-sight. Right: All rows correspond to an inclination of $i = 30^{\circ}$. The first three rows of stellar velocity maps are with $ \varphi = 30^{\circ} - 90^{\circ}$ in $30^{\circ}$ intervals, with $\gamma = 0^{\circ}$. The third stellar velocity map shows the galaxy with the bar near end-on. The final two rows correspond to $i = 30^{\circ}$, $ \varphi = 0^{\circ}$ and $\gamma = 60^{\circ} - 90^{\circ}$ in $30^{\circ}$ intervals and depict only a rotation in the viewing plane. All model particles are binned, thus the width of each map is approximately 40 kpc.
        \vspace{2em}}
        \label{figure:orientations_horizontal}
    \end{figure*}

    Figure \ref{figure:orientations_horizontal} demonstrates the buckling signal for an inclined view of the model\textcolor{red}{, with the mean stellar line-of-sight velocity plotted in each bin}. We find that the buckling signal is detectable through the kinematic signal, by eye, from any $\varphi$ as long as the inclination is fairly low ($i \lesssim 50^{\circ}$). This effect can be also observed in Figure \ref{figure:quadmoment} through the quadrupole moment. Buckling is also detectable through a C-shaped signal, when the bar is viewed end-on. This signal, however, is less useful when comparing to observational data since an end-on bar is difficult to identify through photometry.

    \subsection{Buckling Signal Amplification with Unsharp Masking
    }
    \label{sec:sig_amp}

	We use digital unsharp masking to amplify the buckling signal. Unsharp masking is a linear image processing technique developed initially for photography, but is often utilized in astronomical image processing (e.g. \citealt{2006MNRAS.370..753B}) as well as in simulation image processing (\citealt{2005MNRAS.358.1477A}, Fig. 6).
	
	For each pixel $p$ in the map, we take the average value (with 5$\sigma$ outliers removed) of the velocity signal of all other pixels within a circular aperture of specified radius from $p$. We exclude pixels that are partially inside the aperture. Then, we subtract this average value from the value at the central pixel $p$. Unsharp masking enhances local extrema: using smaller radii for averaging enhances finer detail, and using larger radii can enhance larger features, such as spiral arms. The results from filtering with larger radii are more visually similar to the original map. Subtracting the median instead of the mean generates similar results. The results are also similar to other techniques, such as fitting the velocity within the aperture as a linear function of position ($V_r = V_{r,0} + m(x-x_0) + n(y-y_0)$) and then replacing the central pixel value with the fit residual at that point. 
	
\begin{figure}
        \centering
        \includegraphics[width=1.0\linewidth]{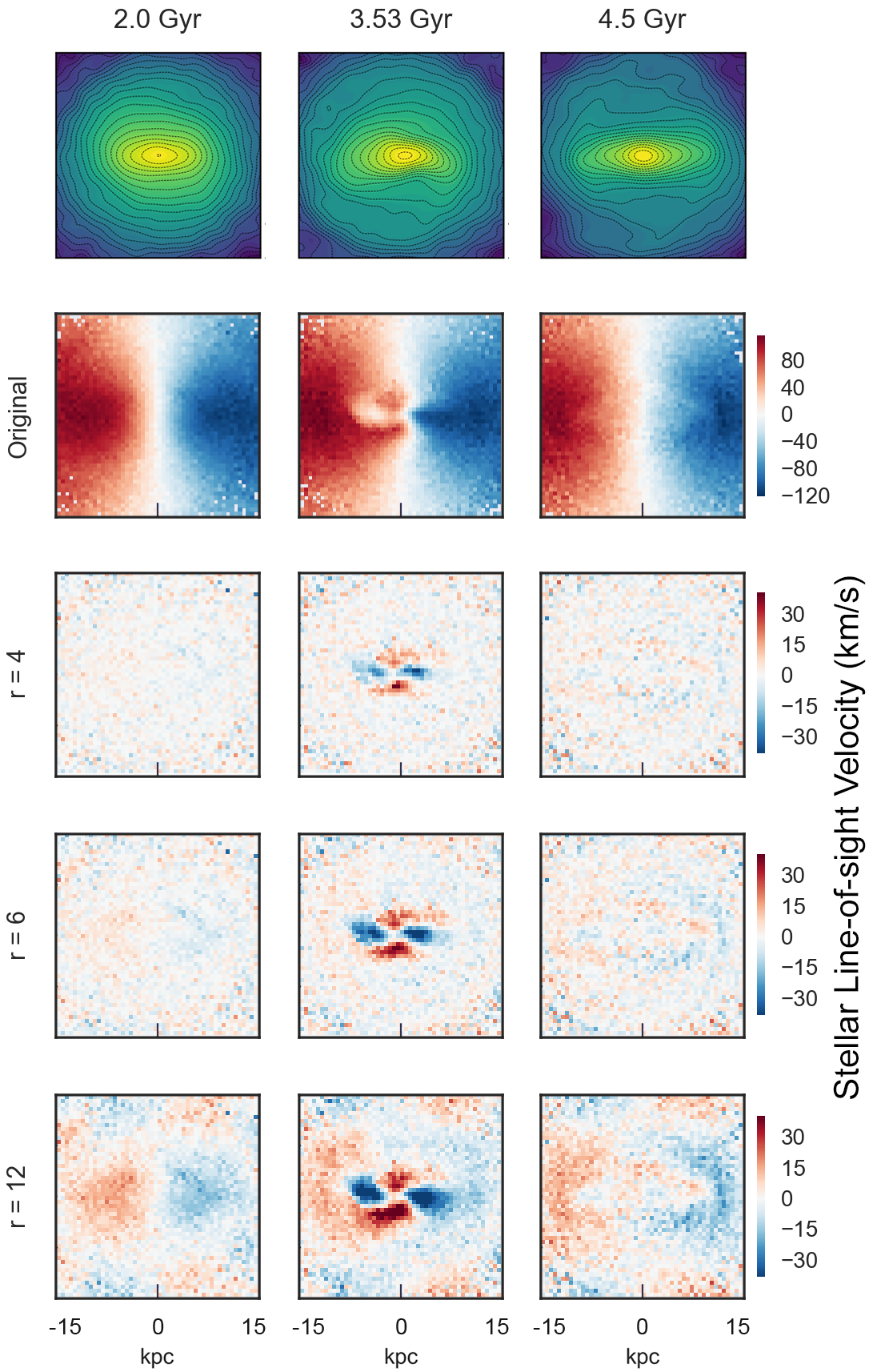}
        \caption{Unsharp masking on simulated stellar velocity maps. Top row shows contour plots of the density for $i = 30^{\circ}$, $\varphi = 0^{\circ}$,  $\gamma = 0^{\circ}$. From left to right, the columns depict the pre-buckling 2.0 Gyr snapshot, the buckling 3.53 Gyr snapshot, and the post-buckling 4.5 Gyr snapshot. The next four rows depict the stellar velocities, with the latter three rows depicting unsharp filtered data with a radius of r = 4, r = 6, and r = 12 pixels, respectively. The bar measures 28.5 pixels in these maps.}
        \label{fig:unsharp_masking}
    \end{figure}

	The effects of unsharp masking on a simulated stellar velocity map are displayed in Figure \ref{fig:unsharp_masking}, illustrating that this technique helps visually enhance the buckling signal. For this model, an unsharp-masking radius of six pixels works \textcolor{red}{best. Given our binning of 0.4 kpc/pixel, this length corresponds to 2.4 kpc.}
    
	\section{Data \& Analysis} \label{sec:Analysis}
	
	\subsection{Sample Selection}
We \textcolor{red}{now proceed to comparing our simulated buckling signal to observational data. Our parent sample is a large sample of barred galaxies from the MaNGA survey \citep{2015ApJ...798....7B}. MaNGA is part of the fourth-generation Sloan Digital Sky Survey (SDSS-IV, \citealt{2017AJ....154...28B}) with data taken at the SDSS 2.5 m telescope located at the Apache Point Observatory in New Mexico \citep{2006AJ....131.2332G}. MaNGA is an ongoing project to map 10,000 galaxies with an integral field unit (IFU) system \citep{2015AJ....149...77D} which feeds into a multi-object spectrograph \citep{2013AJ....146...32S}. As a large IFU survey of galaxies, it is particularly well-suited to our purposes of identifying and testing kinematic signatures of bar buckling. }

\textcolor{red}{The MaNGA survey design is described
by \citet{2016AJ....152..197Y}, with the target selection for the survey presented by \citet{2017AJ....154...86W} and with the observing strategy detailed by \citet{2015AJ....150...19L}. The data reduction pipeline and the spectrophotometric calibration are outlined by \citet{2016AJ....152...83L} and  \citet{2016AJ....151....8Y}. To access MaNGA data, we used Marvin \citep{Cherinka_2019}, a web interface and an application programming interface (API) that allow for analysis of IFU data. Here we make use of products from the MaNGA Product Release 8 (MPL-8).}

We use a barred subset of the sample\textcolor{red}{, with bar lengths and angles as measured by \citet{fraser-mckelvie_merrifield_aragon-salamanca_masters_2019}, relying on the method detailed in \citet{2012ApJ...757...60K}.} Bar classifications are obtained from Galaxy Zoo 2 (GZ2, \citealt{2013MNRAS.435.2835W}). Galaxies in this sample satisfy a debiased bar likelihood \texttt{p\_bar\_weighted > 0.5} and \texttt{p\_not\_edgeon > 0.5} in GZ2, and a $b/a$ -- optical minor to major axis ratio of the outer disk -- of greater than 0.6, which roughly corresponds to inclinations of less than 55$^{\circ}$. We use $b/a$ for elliptical apertures, obtained by taking the axis ratio using Stokes parameters at 90\% light radius, in the NASA-Sloan Atlas (NSA, \citealt{2011AJ....142...31B}), a catalog of images and parameters of local galaxies from SDSS imaging. In this study, we use the version available for MaNGA, \texttt{v1\_0\_1}. The sample contains 434 galaxies with measured bar lengths and angles.

\subsection{Quadrupole Moment Analysis}
\label{sec:qmomentanalysis}

We use stellar velocity maps that are Voronoi binned to a signal-to-noise ratio of 10 for all of these galaxies, along with unsharp filtered images using radii of 4, 6, and 8 pixels. \textcolor{red}{Then, we calculate $q$ for the unsharp filtered maps by using the stellar velocity per bin rather than per model particle, as for the snapshots.} We then use the lowest $q$ across the three unsharp filtered maps to rank galaxies based on a consistently high quadrupole moment.

\begin{figure*}
    \centering
    \includegraphics[width=1.0\linewidth]{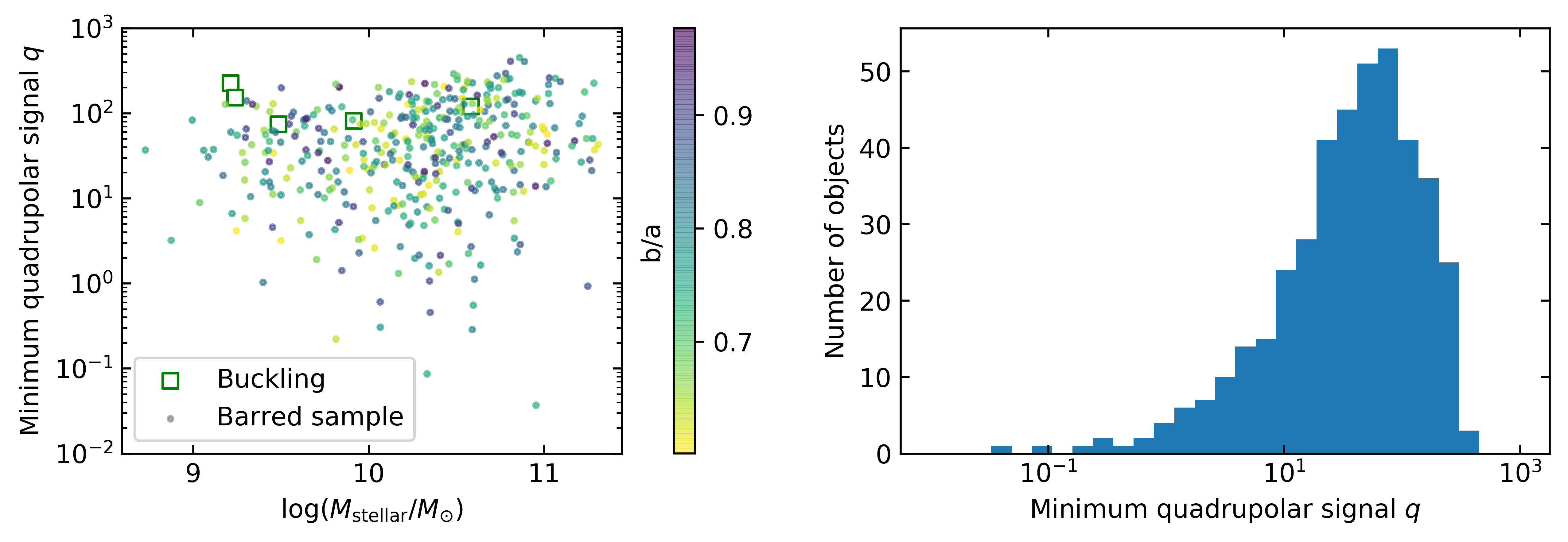}
    \caption{Left: Lowest quadrupole moment ($q$) across r=4, r=6, and r=8 unsharp filtering, plotted against log stellar mass, for observed galaxy \textcolor{red}{stellar velocity} maps. \textcolor{red}{Colorbar corresponds to axial ratio $b/a$.} Buckling candidates indicated \textcolor{red}{with an empty square}. Right: Histogram of all 434 galaxies binned by \textcolor{red}{minimum} $q$. \vspace{2em}}
    \label{fig:quadhisto}
\end{figure*}

After visually inspecting the top 100 galaxies, we identify sixteen preliminary high-quadrupole moment galaxies with velocity maps resembling a buckling signal: a central disturbance in the stellar line-of-sight velocities, resembling a quadrupole or a bent C-shaped asymmetry, such as in Figure \ref{figure:orientations_horizontal}. A distribution of quadrupole moments and their dependence on mass \textcolor{red}{and $b/a$} is displayed in Figure \ref{fig:quadhisto}. 

We also considered normalizing the quadrupole moment by the dipole moment \textcolor{red}{(as in Figure \ref{figure:quadmoment}, inset)} and used the metric $q/d = \sum_{n = 0}^{N} [V_{\rm{los, }n} \cos(2(\alpha_n-\alpha_0))]/\sum_{n = 0}^{N}[V_{\rm{los, }n}\cos(\alpha_n-\phi)]$, where $\phi$ is varied from $0^\circ$ to $180^\circ$. We then normalize by the largest dipole moment. Using $q/d$ as a metric yields the same sixteen initial buckling candidates as well as qualitatively similar results.

\subsection{Buckling verification and orientation fitting}
\label{optimization}
\textcolor{red}{As a consistency check for buckling candidate identification, we directly compare the observational stellar velocity} maps to the model. We select 16 preliminary buckling candidates identified from their large $q$ values. One additional candidate is chosen by visual inspection from the remaining galaxies in MPL-8 that are not in \citet{fraser-mckelvie_merrifield_aragon-salamanca_masters_2019}. 

We rotate the simulated galaxy and generate stellar velocity maps that match observed maps, optimizing the viewing parameters. A correlation coefficient between stellar velocities at each pixel was obtained for each simulated-observed pair, and we minimize this value. For pixels that are binned together to satisfy a signal-to-noise ratio of 10, we only take into account one pixel in each bin. The mask for the observational map is then identically applied to the simulation map. Since the calculation is computationally expensive, we select just 26 additional barred galaxies with $b/a$ > 0.6 at random as a control sample for comparison.

We fit the stellar velocity maps of the 16 candidates (and of the 26 galaxies in the control sample) to both the buckling and post-buckling simulation snapshots. We choose to fit to the post-buckling snapshot instead of the pre-buckling snapshot, since the observed galaxies were selected on the basis of possessing a strong bar. Therefore, if the observed galaxy in question is not buckling, it \textcolor{red}{more likely  is in the longer lasting secular evolution stage (\citealt{2013seg..book..305A}, for a
review) and thus has already buckled.}

We first use a brute force minimization algorithm (\texttt{scipy.optimize.brute}), which computes a cost function for a multidimensional grid of parameters, detailed in Table \ref{table:brute_parameters}. The parameters include the viewing angles, size scale, and velocity scale of the simulation snapshot. The cost function is the median of the pixel-wise absolute difference (MAD) between the maps, and we seek to minimize this value. The centers of the observed and simulated maps are aligned and fixed based on the center of mass of the $r$-band flux and the center of the map, respectively. The mask for each observed map is also applied to each generated simulation snapshot map for all iterations.

To minimize the cost function, we vary five parameters: the three angles associated with the galaxy's orientation ($i$, $\varphi$, $\gamma$) as well as scale factors for the bar length and the stellar velocity amplitude\footnote{\textcolor{red}{Since re-scaling the spatial extent and the velocity amplitude affects the galaxy's mass, dynamics, and relevant timescales, we use the N-body simulation only to compare and match the pattern in stellar velocities, and not to infer timescales.}}. This method allows us to find the general location of the global minimum in our parameter space, and to reduce degeneracies. The set of parameters that provides the lowest value is recorded. The bar length parameter is fixed based on our initial guess\textcolor{red}{, though we find that the final fit parameters are robust to as much as a 50\% difference in the bar length guess. }

\begin{table}
\begin{center}
\begin{tabular}{|l|l|l|}
\hline
Parameter                  & Range                  & Step \\ \hline
$i$ (degrees)              & Init. guess $\pm$ 20 & 10             \\ \hline
$\varphi$ (degrees)        & [-180, 180]            & 10            \\ \hline
$\gamma$ (degrees)         & [-180, 180]            & 10            \\ \hline
Spatial resolution scale (pixels) & Init. guess        & ---          \\ \hline
Velocity amplitude scale & [0.5, 2.0]             & 0.5           \\ \hline
\end{tabular}
\end{center}
\caption{Fitting parameters and grid details for the brute-force optimization. The rightmost column details the step size in the parameter range. Velocity amplitude is re-scaled before fine-tuning. The spatial resolution scale is fixed for the brute force optimization and allowed to vary slightly when fine-tuning.}
\label{table:brute_parameters}
\end{table}

After brute force optimization, we take the best result and rescale the simulation velocity amplitude. The default simulation velocity amplitude is roughly 200 km/s. We check scale values from 0.4 to 2.5, in increments of 0.05, and then record the parameters that yield the lowest MAD. We then we use a downhill Nelder-Mead simplex algorithm \citep{10.1093/comjnl/7.4.308, Gao2012} to polish the result, with an initial guess given by the brute force optimization result. Each iteration takes the MAD between the simulated and observed maps (with additional soft constraints) as a cost function, and seeks to minimize this value. In the minimization process, a simplex of $(n+1)$ vertices is generated for $n$ parameters, and the cost function is evaluated at each vertex. Each vertex slightly varies one parameter, and the values of the cost-function are compared between all vertices. The vertex with the highest value (i.e. the worst fit) is removed, and a new vertex replaces it. Thus, the simplex contracts towards the final minimum, and adapts to the local landscape. We run the optimization for 100 iterations before stopping at the parameters that yield the lowest MAD. 

The entire fitting procedure is performed for each observed galaxy twice: once to the buckling snapshot, and once to the post-buckling snapshot. Both steps of fitting are performed with soft constraints to ensure that the orientation and radial extent of the simulation do not differ significantly from that of the observational data. 

\subsection{Constraints} \label{sec:Constraints}
Fitting stellar velocities places a primary constraint on the galaxy orientation, but we require the inclination and bar angle of the model view and the observed galaxy to be consistent as well. Our three priors are inclination, bar position angle, and bar length, and are described below. Below we elaborate on the soft constraints that we implement.

\paragraph{Inclination}\label{par:Inclination}
Inclination of each observed target is informed through the axial ratio $b/a$. A ratio of $b/a = 1$ indicates that a galaxy is near face-on, and $b/a$ decreases as the inclination increases towards an edge-on view. To obtain the inclination, we use the relation given by \citet{2010MNRAS.404..792M}:
\begin{equation}\label{equation:ba_inclination}
    \cos^2(i) = \frac{(b/a)^2-p^2}{1-p^2}  
\end{equation}
where $b/a$ is \textcolor{red}{obtained from the NSA (\texttt{BA\_90}, or in MaNGA, \texttt{elpetro\_ba}),} and $p$ is the intrinsic axial ratio. We use the $b/a$ from an elliptical Petrosian fit; the Petrosian radius is defined here to be \textcolor{red}{the radius at which} the $r$-band surface brightness in a local annulus is 20\% of the mean $r$-band surface brightness within that radius. Further details are available on the SDSS webpage\footnote{\url{https://www.sdss.org/dr16/manga/manga-target-selection/nsa}}. To account for the non-zero thickness of the disk, we use $p = 0.12 + 0.10 f_{\mathrm{DeV}}$ \citep{2010MNRAS.404..792M}\textcolor{red}{, where }$f_{\mathrm{DeV}}$ is the fraction of light that can be fit by a de Vaucouleurs profile. We use the $r$-band $f_{\mathrm{DeV}}$ from SDSS since it has the highest signal-to-noise ratio. A linearly increasing penalty is given for a simulation inclination angle that deviates from the observed inclination angle by more than 15 degrees. 
    
    \paragraph{Bar angle}\label{par:Bar_angle}

    For observational target 8947-1901, and a number of the control galaxies, the bar position angle is estimated by using an average of angles generated from ellipse fits of isophotes using \texttt{astropy Photutils} \citep{larry_bradley_2019_2533376}. We use the mean $r$-band flux per pixel, and define bar angle counterclockwise from East; a horizontal bar has a bar position angle of $0^\circ$. For the remaining targets, we use the bar angles measured by \cite{fraser-mckelvie_merrifield_aragon-salamanca_masters_2019}. Elsewhere, uncertainty in bar position angle measurement is one of the key uncertainties in determining other galaxy properties such as bar pattern speed \citep{2003MNRAS.342.1194D,2019MNRAS.482.1733G}. Hence, a linearly increasing penalty is given if the fit bar angle for the observational map and the calculated bar angle of the simulation map differ by more than $10^\circ$.
    
    \paragraph{Bar length}\label{par:bar_len}
    We use the length of the bar as a parameter for aligning the spatial resolution scale of the observational and simulation maps. MaNGA stellar velocity map bins are 0.5 arcseconds per pixel; we alter the bin size for the simulated galaxy to ensure that the measured length of the bar in the observed map matches the projected length of the bar in the simulated map, in pixels. The bin size parameter is first fixed in the brute force optimization, and then allowed to vary slightly when fine tuning parameters. A linearly increasing penalty is given if the bar length deviates by more than 4 pixels from the initial guess, which places a strict constraint on bar length. Observed bar lengths are obtained from \cite{fraser-mckelvie_merrifield_aragon-salamanca_masters_2019} and \cite{2011MNRAS.415.3627H}, and averaged when available in both samples. We remeasure bars with length guesses that appear to be incorrect by significantly more than 4 pixels, to ensure closer fit results.

   \subsection{Results} 
   \label{sec:results}
   
    \begin{figure*}
        \centering
        \includegraphics[width=0.7\linewidth]{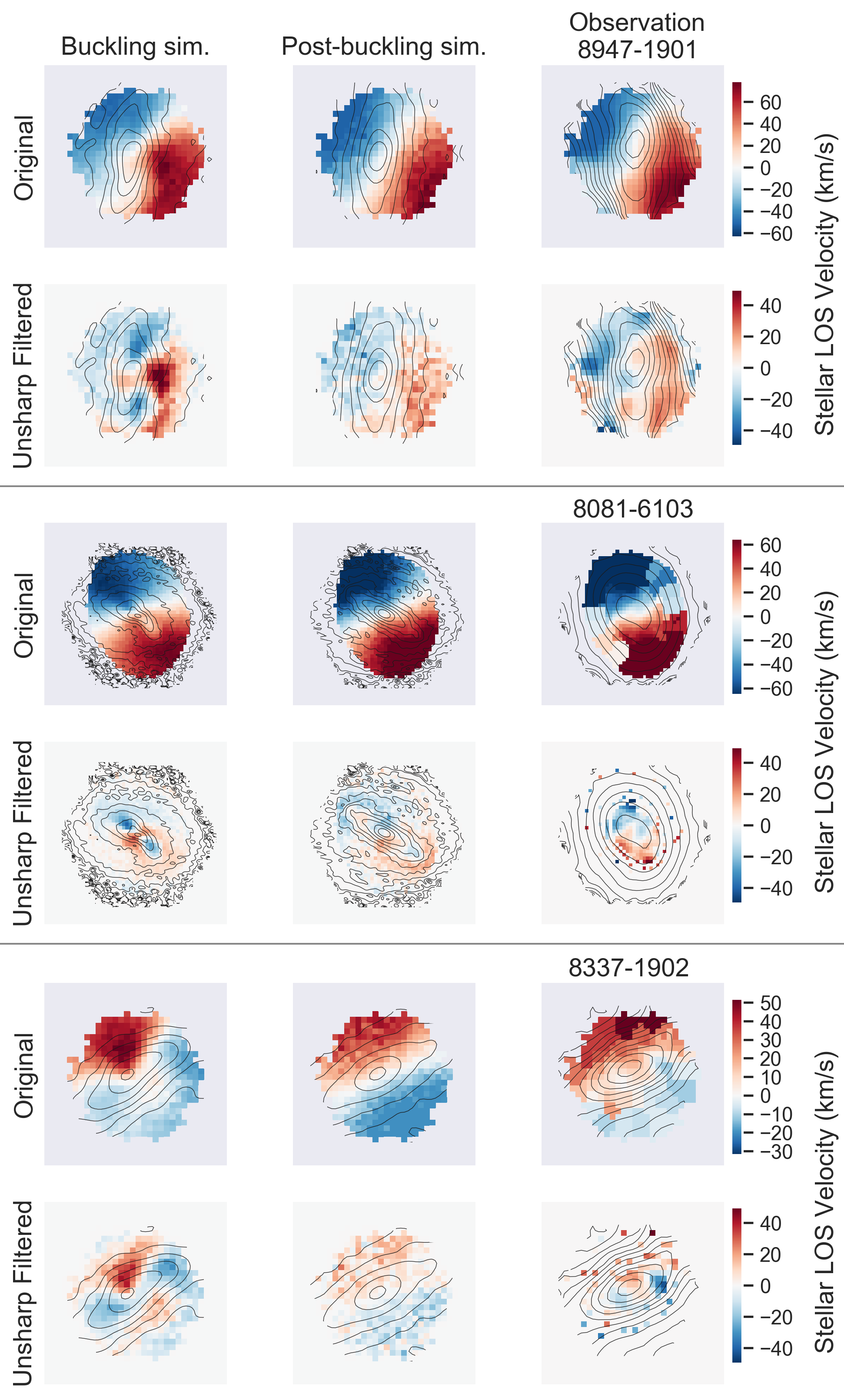}
        \caption{Stellar velocity maps for three of the five buckling candidates. In each panel, the first and second columns correspond to buckling and post-buckling simulation snapshots, respectively. The third column corresponds to the observed galaxy. The first row depicts stellar velocity maps, and the second row depicts the unsharp filtered stellar velocity maps, with an aperture of radius 6 pixels. \textcolor{red}{Logarithmically spaced isodensity contours are overplotted, and are estimated from the particle count in each bin and the mean $r$-band flux, for the simulated and observed galaxies, respectively.} Simulated stellar velocity maps are generated using the best-fit parameters (see Table \ref{table:fit}) and the optimization algorithm detailed in Section \ref{optimization}. \vspace{1em}}
        \label{fig:buckling_1}
    \end{figure*}
    \begin{figure*}
        \centering
        \includegraphics[width=0.7\linewidth]{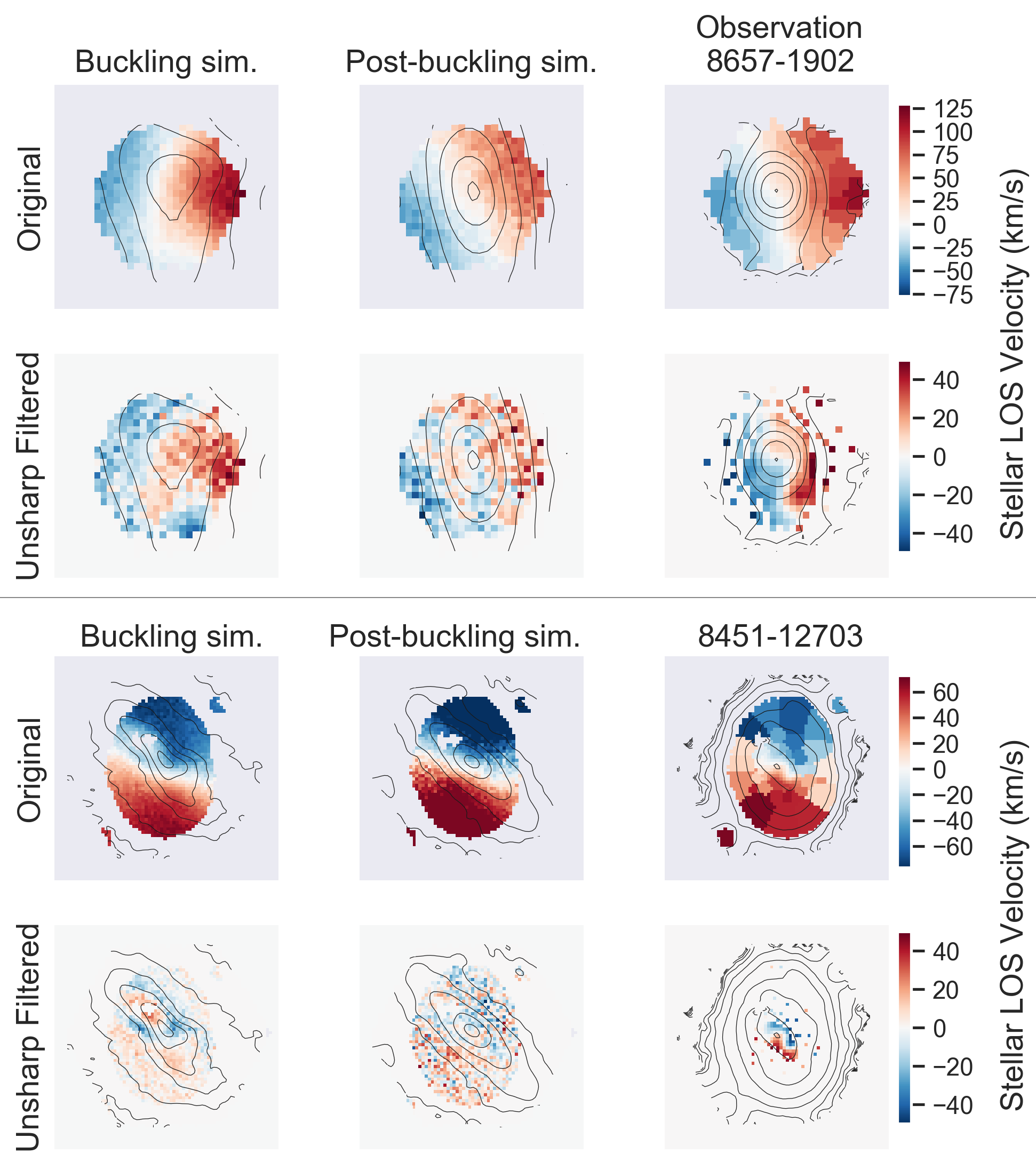}
        \caption{Stellar velocity maps for the remaining two of the five buckling candidates, with plot format identical to Figure \ref{fig:buckling_1}.\vspace{1em}}
        \label{fig:buckling_2}
    \end{figure*}

\begin{table*}[t]
\begin{center}
\begin{tabular}{@{}llllll@{}}
\toprule
Name & Plate-IFU & $i$ ($^{\circ}$) & ra & dec & z \\ \midrule
LEDA 1160737 & 8081-6103 & 46 & 48.06547714 & 0.240451563 & 0.038046 \\
MCG+06-31-100 & 8337-1902 & 19 & 214.9040217 & 38.23424387 & 0.020877 \\
LEDA 32858 & 8451-12703 & 51 & 164.0289 & 43.15656828 & 0.037901 \\
MCG+00-02-097 & 8657-1902 & 19 & 9.564853953 & -0.967405561 & 0.067841 \\
2MASX J11253210+5110494 & 8947-1901 & 36 & 171.383699 & 51.1804088 & 0.027127 \\
NGC 4569, M90 & N/A & 69 & 189.207567 & 13.162869 & -0.000784 \\
NGC 3227 & N/A & 63 & 155.87735 & 9.865197 & 0.003650 \\
ESO 506-4 & N/A & 61 & 294.325537 & 38.208011 & 0.013223 \\\bottomrule
\end{tabular}
\end{center}
\caption{Brief summaries of buckling candidates. The first five are part of the MaNGA sample, and identified in this paper, the next two have been identified in \cite{2016ApJ...825L..30E}, \textcolor{red}{and the last one in \cite{Li_2017}}. Inclinations for the first five rows taken from the results of the fitting procedure; the rest are from the two papers mentioned above. For NGC 3227, the inclination is obtained from \cite{Fischer_2013}. Redshifts were obtained from the NSA, and for ESO 506-4, from \cite{2011MNRAS.416.2840L}.}
\label{table:allcandidates}
\end{table*}

    In this section, we present the results of the comparison between observations and simulations. \textcolor{red}{The buckling candidates displayed possess a high quadrupole moment consistent with simulation buckling, and are shown alongside the simulated stellar velocity maps.}
    
   \paragraph{Buckling Candidates}
    Out of the 16 initial buckling candidates, we select 5 final candidates with low buckling fit median absolute deviation, and well-aligned stellar velocity maps. The stellar velocity maps and the fits of the five buckling candidates are displayed in Figure\textcolor{red}{s \ref{fig:buckling_1} and \ref{fig:buckling_2}.} The fitting procedure \textcolor{red}{mitigates the effect of} false positives: 11 candidates from the original buckling sample of 16 are removed. Although these 11 galaxies possess a quadrupolar signal near the center of the line-of-sight stellar velocity maps, the shape and size of the signal are inconsistent with bar angle and size constraints. Therefore, we reject these candidates from the initial buckling sample. Additionally, we find that one of these high quadrupole-moment galaxies (MaNGA plate-IFU 8722-3702) has been mistakenly identified as having a bar. The GZ2 redshift-debiased sample generally identifies bars accurately, but in this case, the high redshift (z $\approx$ 0.11) resulted in a high \texttt{p\_bar\_debiased = 1}. Yet, the original bar likelihood from volunteer identifications was low (\texttt{p\_bar\_weighted = 0.33}). Thus, we have disregarded 8722-3702 and have excluded it from Figure \ref{fig:fitres}.

    Final fit information for each parameter is shown in Table \ref{table:fit}. Priors and details are are shown in Table \ref{table:details}. For many of these candidates, the non-buckling MAD is \textcolor{red}{yet} lower than the buckling fit MAD. We believe that this result is due to the difference in intensity of buckling signal -- these observed galaxies may be in a different stage of buckling from the simulation snapshot. Figure \ref{fig:fitres} provides a comparison of the buckling best fit MAD and the non-buckling best fit MAD, along with the 26 control galaxies. \textcolor{red}{Galaxies labeled as ``unlikely buckling" are less likely to be buckling due to a lower quadrupole moment; regardless, we cannot exclude the possibility.} Galaxies that are visually identified as non-buckling are also labeled.
    
    \textcolor{red}{Furthermore, we recall that the N-body model we used exhibited a significant in-plane bending of the bar. We note that a number of the buckling candidates show this effect as well; this can be observed from the isodensities in Figures \ref{fig:buckling_1} and \ref{fig:buckling_2}. One bar is strongly asymmetric (8947-1901), three are less so (8081-6103; 8337-1902; 8451-12703) and only one is definitely not (8657-19020). As our analysis only involved comparing kinematics and bar angle, it is promising that the in-plane bending, a predominantly photometric feature, appears to be consistent between simulations and observations as well.}
    
    The candidates found from our analysis brings the total number of buckling candidates \textcolor{red}{characterized to eight; two} have been discovered in \citet{2016ApJ...825L..30E}\textcolor{red}{, and one more probable buckling candidate has been characterized in \citet{Li_2017}. Details for all eight} candidates are collected in Table \ref{table:allcandidates}.
    
   \paragraph{Non-buckling comparison galaxies}
   
\begin{figure*}
    \centering
    \includegraphics[width=0.7\linewidth]{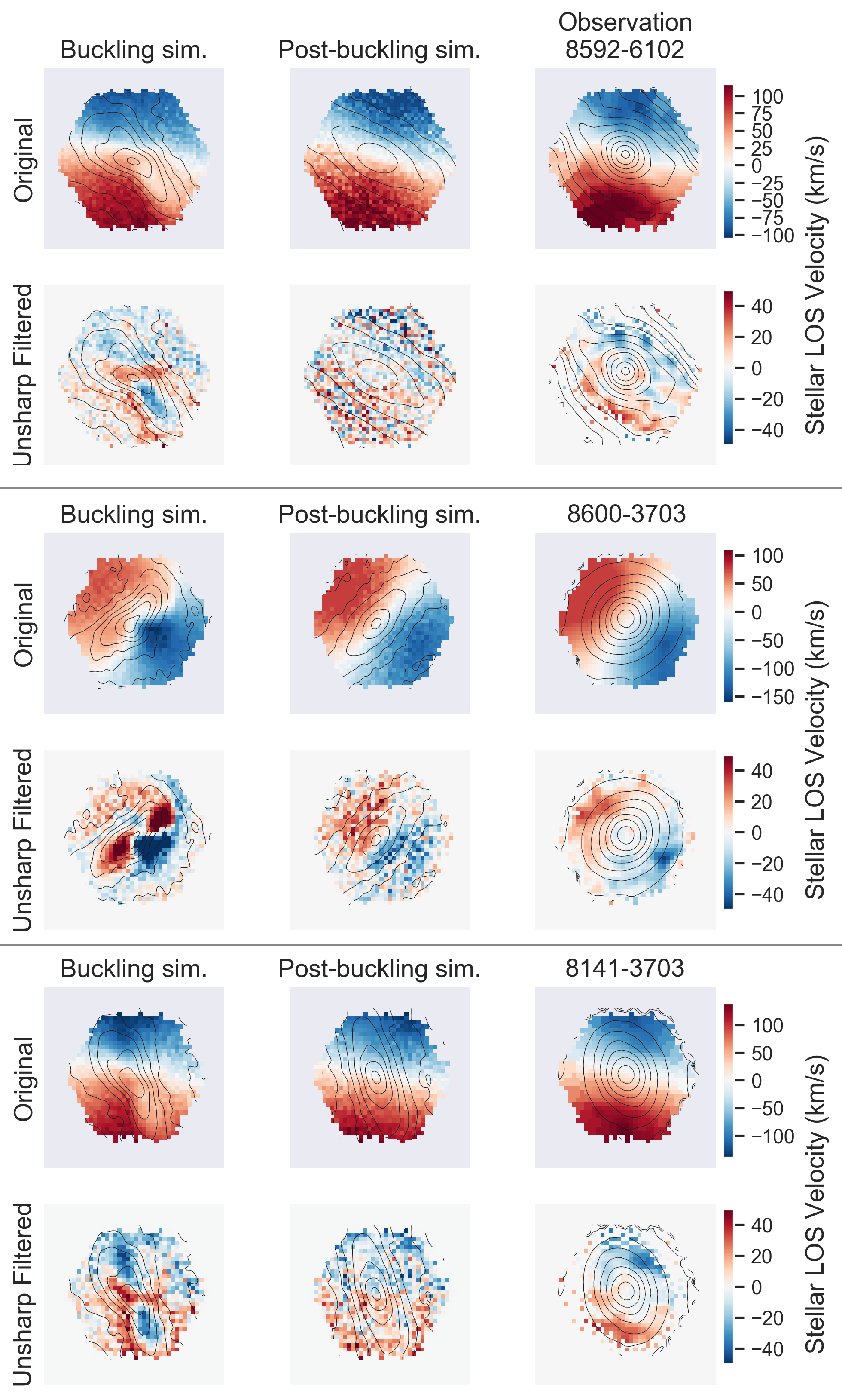}
    \caption{Stellar velocity maps of three galaxies deemed to be not buckling, with plot format identical to Figure \ref{fig:buckling_1}.}
    \label{fig:post-buckling_candidates}
    \vspace{2em}
    \end{figure*}

\begin{figure}
    \centering
    \includegraphics[width=\linewidth]{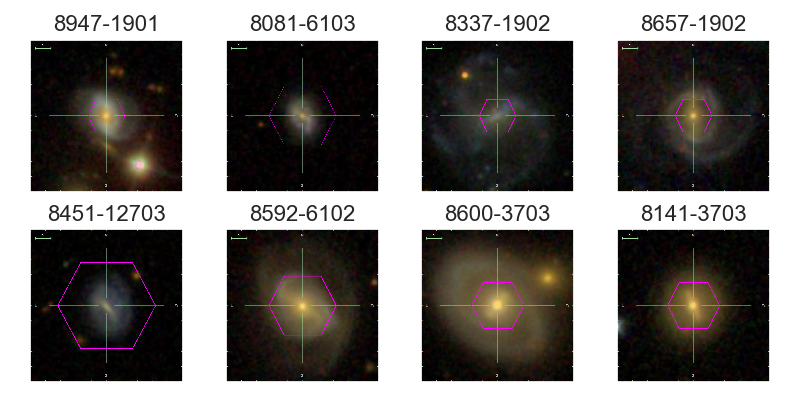}
    \caption{\textcolor{red}{Optical images of the galaxies described in Figures \ref{fig:buckling_1}-\ref{fig:post-buckling_candidates}. The objects to the lower right of 8947-1901 and the upper left of 8337-1902 are foreground stars; the object to the upper right of 8600-3703 is a non-interacting higher redshift galaxy.}}
    \label{fig:optimg}
    \vspace{2em}
    \end{figure}

\begin{figure}
        \centering
        \includegraphics[width=1.0\linewidth]{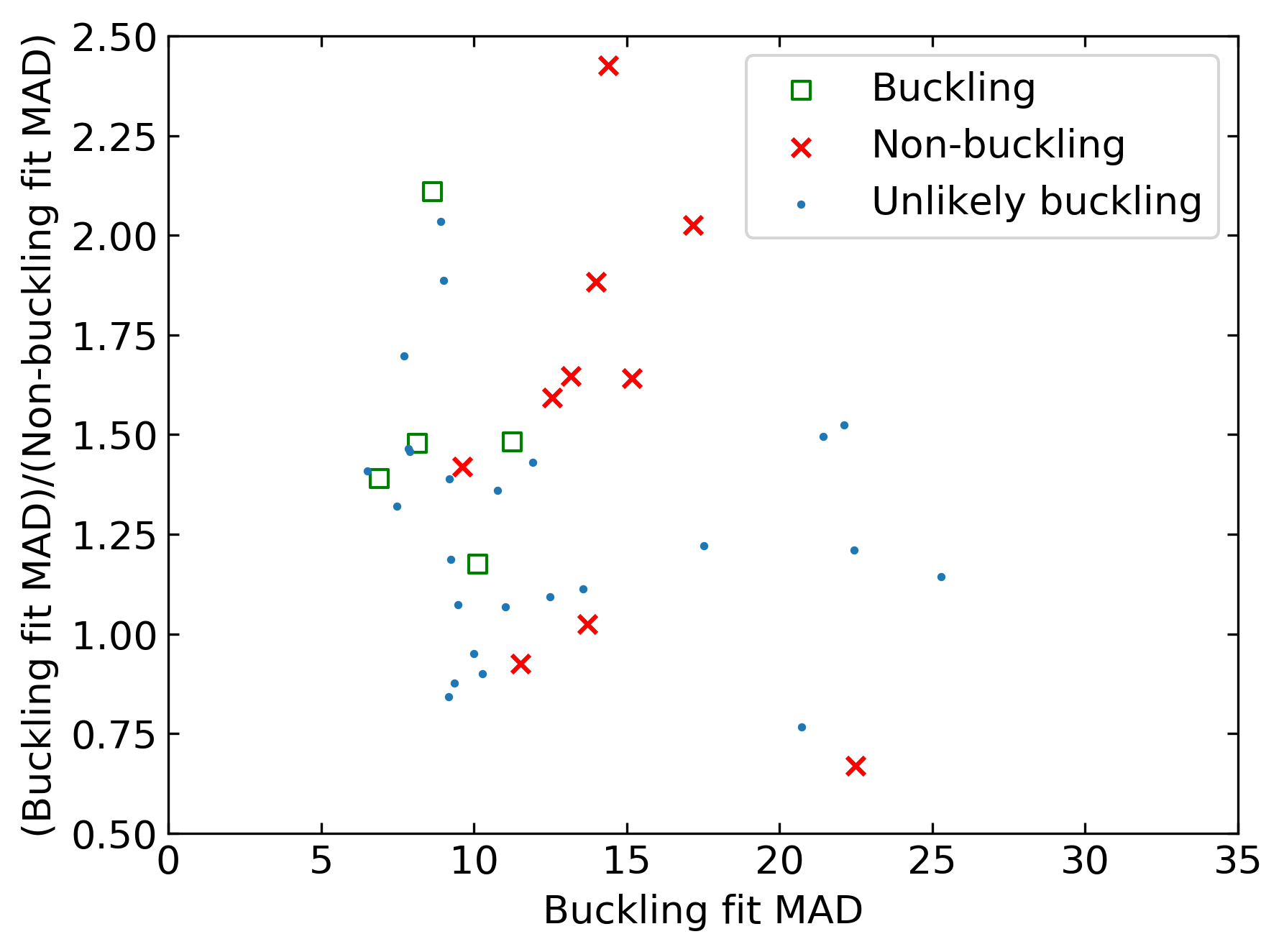}
        \caption{Median of the pixel-wise absolute difference (MAD) between the observed and simulated stellar line-of-sight velocity maps, for each galaxy in Table \ref{table:details}.}
        \label{fig:fitres}
    \end{figure}

   We provide a sample of galaxies which we believe are not buckling, for comparison. We again fit both the 4.5 Gyr post-buckling model and the 3.53 Gyr buckling model to the observed stellar velocity maps. Three sample non-buckling comparison galaxies are shown in Figure \ref{fig:post-buckling_candidates}. \textcolor{red}{The optical images for all of the galaxies plotted are shown in Figure \ref{fig:optimg}.}
 
   \subsection{Gas masses}\textcolor{red}{
   N-body simulations predict that buckling should be affected by the gas fraction. To test these predictions, we obtain }HI masses from the Arecibo Legacy Fast Arecibo L-band Feed Array (ALFALFA) survey \citep{2011AJ....142..170H, 2018ApJ...861...49H} as well as from HI-MaNGA follow-up observations at the Robert C. Bryd Green Bank Telescope \citep{2019MNRAS.488.3396M}.  
    
    HI mass data are matched to the MaNGA Product Release 8 barred sample, and yield a total of 124 galaxies. (10 galaxies are removed from the sample because they are edge-on.) We correct for self-absorption using the optical axial ratio, with $M_{\textrm{HI, }c} = c M_{\textrm{HI}}$, where $c = (b/a)^{-0.12}$ \citep{1994AJ....107.2036G}. Stellar masses are obtained from the NSA, via a K-correction fit for elliptical Petrosian fluxes.

     HI masses are derived from fluxes as follows: 
     \begin{equation}
         M_{\textrm{HI}}/M_{\odot} = 2.356 \times 10^5 \Big( \frac{D}{\textrm{Mpc}} \Big)^2 \Big( \frac{F_{\textrm{HI}}}{\textrm{Jy km/s}} \Big)
     \end{equation}
     
    \noindent $F_{HI}$ is the HI flux, and we assume that $D = c z H_0$ where $z$ is the optical redshift of each MaNGA galaxy in the NSA. Error for HI masses is propagated via the error on the HI flux: 
     \begin{equation}
         F_{\textrm{HI, error}} = rms \sqrt{\Delta\nu W}
     \end{equation}
     \noindent where rms is the root mean square noise, $\Delta \nu$ is the channel resolution, $10$ km s$^{-1}$, and $W$ is the width of the profile, which we approximate as $1.2 W_{\mathrm{P20}}$. $W_{\mathrm{P20}}$ is the width of the HI line measured at 20\% of the peak of the HI line. Typical errors are around $\delta  (M_{\textrm{HI}}/M_{\odot}) / (M_{\textrm{HI}}/M_{\odot}) = 0.08$.

\begin{figure}
        \centering
        \includegraphics[width=1.0\linewidth]{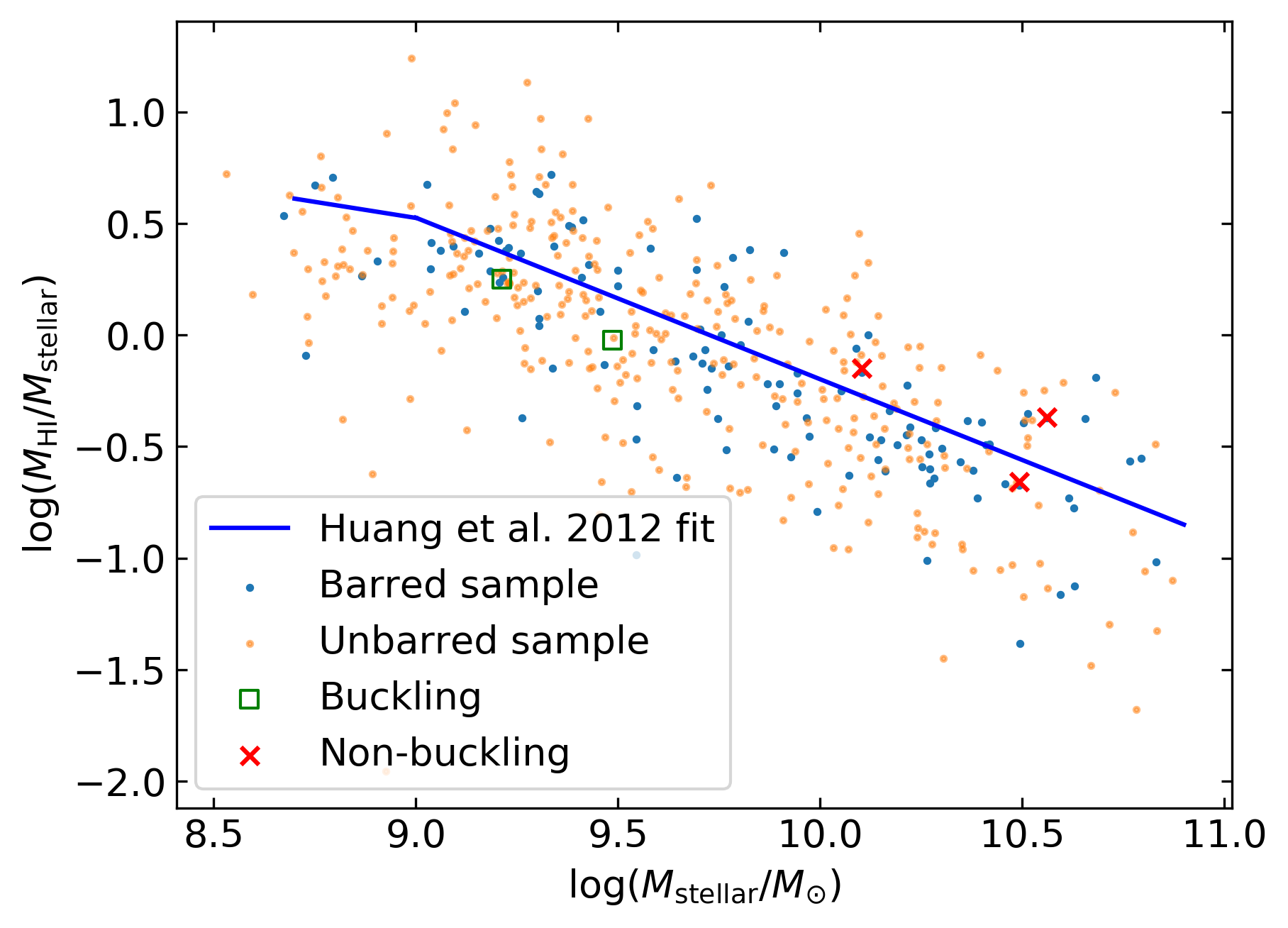}
        \caption{HI mass to stellar mass ratio, versus stellar mass, for all MaNGA galaxies cross-referenced with ALFALFA and GBT datasets. (Not all of the galaxies in Table \ref{table:allcandidates} have gas masses available.) \textcolor{red}{Blue and orange: barred and unbarred galaxies, respectively.} Green: buckling candidate galaxies, red: galaxies that are not buckling. Fit line from \citet{2012ApJ...756..113H}.\vspace{1em}}
        \label{fig:gas_masses}
    \end{figure}

    We plot $\log(M_{\rm{HI}}/M_{\rm{stellar}})$ against $\log(M_{\rm{stellar}}/M_{\odot})$ for 124 barred galaxies for which gas masses are available, in Figure \ref{fig:gas_masses}. Buckling and non-buckling galaxies are indicated, and we plot the fit line from \cite{2012ApJ...756..113H}. The two buckling galaxies in the plot are MaNGA Plate-IFUs 8081-6103 and 8451-12703. Non-detections and galaxies for which upper limits are calculated have not been included in Figure \ref{fig:gas_masses}.
    
    \section{Discussion}
    \label{sec:Discussion}

\begin{figure}[h]
    \centering
    \includegraphics[width= 1.0\linewidth]{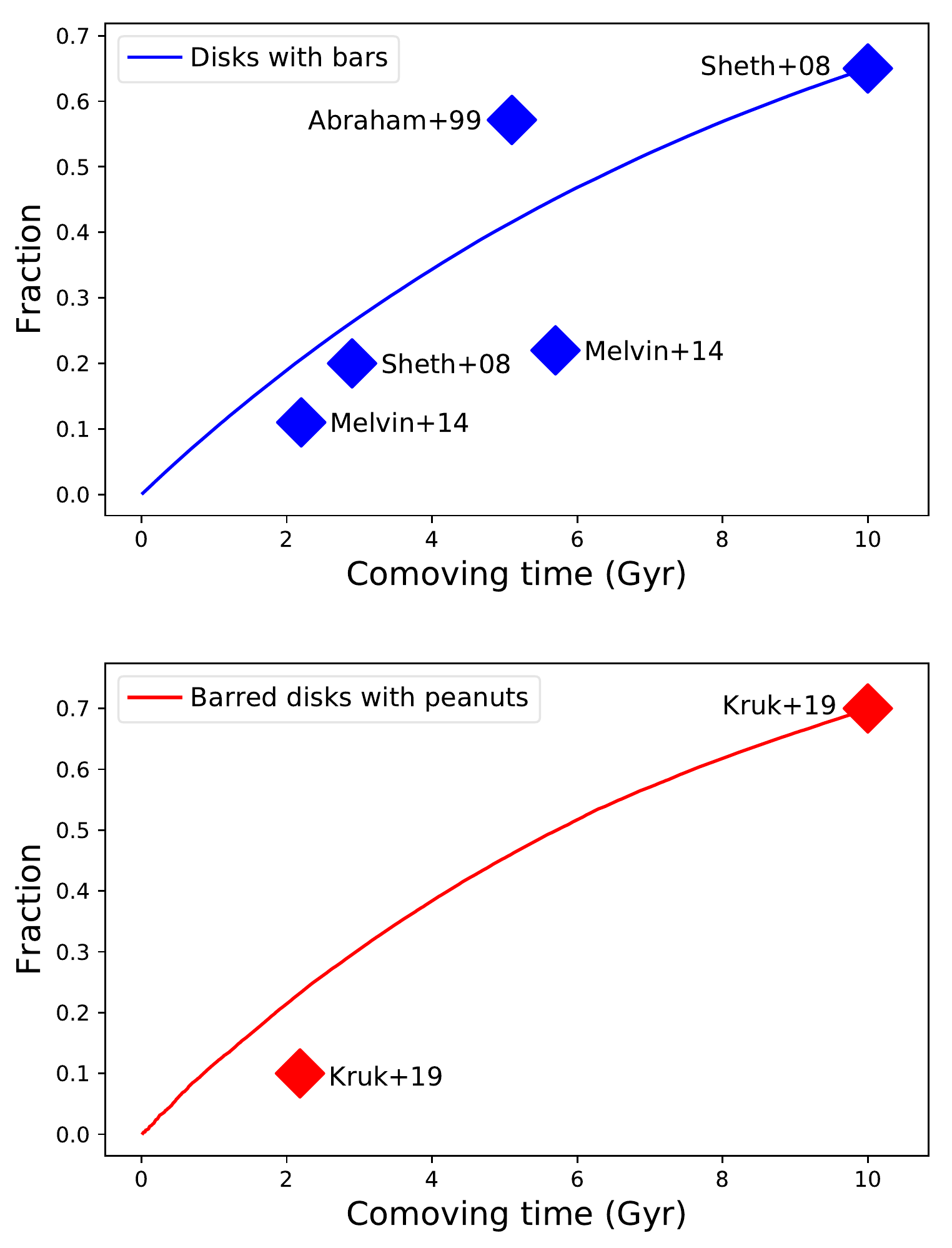}
    \caption{The observational data on the evolution of bar fraction and bar properties with redshift through to the present-day \citep{2008ApJ...675.1141S,2014MNRAS.438.2882M,2017MNRAS.468.2058E,2019MNRAS.490.4721K,10.1046/j.1365-8711.1999.02766.x}, combined with the frequency of observably buckling galaxies measured in this work suggest a characteristic timescale \textcolor{red}{of 9.5 Gyr for bar formation, and 4.0 Gyr for a bar to buckle. The lower bound on the duration of buckling is predicted to be 130 Myr. Top panel: model predictions and data for the bar fraction. Bottom panel: }model predictions and data for the fraction of barred disks that have peanut/X-shaped bars. Here, a comoving time of 10 Gyr denotes the present-day. 
    \vspace{2em}}
    \label{figure:barevolution}
\end{figure}

     \subsection{The fraction of galaxies with bars in the buckling phase}

        Theoretically, the buckling phase is consistently predicted by several different investigations to have a duration of $\sim$0.50-1.00 Gyr \citep{2003MNRAS.346..251O,2004ApJ...613L..29M,2006ApJ...637..214M,2008IAUS..245...93A,2013ApJ...764..123S,2016ASSL..418..391A,2019A&A...629A..52L}. Observationally, \citet{2016ApJ...825L..30E} have identified two disk galaxies whose bars appear to be buckling. These were identified from observations of a sample of 44 barred galaxies with $\log{M/M_{\odot}} \gtrapprox 10.4$ \citep{2017MNRAS.468.2058E}, a sample for which 80\% of the barred galaxies have boxy/peanut bars. They thus measure a frequency of observed buckling in local, high-mass barred galaxies of $f_{\rm{buck}} = 4.5^{+4.3}_{-2.3}\%$.  They then estimated, by incorporating a few assumptions discussed in their Section 4, that high-mass galaxies typically have bars for $\sim$2.2 Gyr prior to buckling taking place, with the buckling instability then lasting $\sim$0.8 Gyr. \textcolor{red}{The results of \citet{Li_2017} are consistent with these timescales, though in both studies, the estimate for the buckling timescale is a lower bound. 
        }
        
        The fraction of objects currently observed to be undergoing buckling places a constraint on the duration of buckling episodes. We examine this process using a semi-analytic model of bar evolution. We fix the percentage of disk galaxies that are barred to be 0\% 10 Gyr ago and 65\% today \citep{2008ApJ...675.1141S,2014MNRAS.438.2882M}, and we assume that the timescale for bar formation, $\tau_{\rm{Bar}}$, is distributed as a Poisson process. We fix the fraction of barred galaxies that are peanuts to be 0\% at the formation of the bar, 70\% today \citep{2019MNRAS.490.4721K}, and we assume that the timescale for a bar to have formed a peanut, $\tau_{\rm{Peanut}}$, is distributed as a Poisson process. Buckling is observable at the end of the buckling process for $\tau_{\rm{Buckling}}$. This timescale \textcolor{red}{relies on} the fact that we identified 5 buckling candidates out of 434 near face-on galaxies, of which no more than $\sim$75\% have been observed with sufficient signal-to-noise and resolution to identify buckling. \textcolor{red}{It should thus be considered as a limiting value.}
        
        We assume that bars are not destroyed, which is supported \textcolor{red}{by a number of high-resolution simulations (see \citetalias{2013MNRAS.429.1949A} for a discussion).} Assuming bars are in general long-lived may indeed be reasonable for non-interacting galaxies, but might not be in cases of strong interactions or of minor mergers \citep{1991dodg.conf..191P,1999ASPC..160..351A,2003MNRAS.341..343B}. Furthermore, interactions can themselves drive bar evolution, including the buckling instability \citep{2019A&A...624A..37L}. Our model does not include the effects of such interactions. The predictions of the bar model, with a comparison to the observational data, are plotted in Figure \ref{figure:barevolution}.

        We find $\tau_{\rm{Bar}} \approx 9.5$ Gyr, $\tau_{\rm{Peanut}} \approx 4.0$ Gyr, and $\tau_{\rm{Buckling}} \approx 130$ Myr. The timescale of 130 Myr is a lower bound on the duration of buckling, as the duration of buckling is necessarily longer than its observability, and some galaxies may form peanuts without buckling \citep{2007ApJ...666..189B,2014MNRAS.437.1284Q}. Given these numbers, the mean age of bars that have already formed is approximately 6 Gyr. Our conversion between cosmological redshifts and lookback times was done using Ned Wright's cosmology calculator \citep{2006PASP..118.1711W}. Given the assumptions of this model, the mean age of bars that exist today is $\sim$5.9 Gyr, with a mean age of bars that have a peanut/X-shape of $\sim$6.7 Gyr, and a mean age of bars that have not yet buckled of $\sim$1.4 Gyr. Of those bars that have buckled, the mean time since the buckling is $\sim$4.4 Gyr. 
        
        \textcolor{red}{
        The amplitude of the quadrupole in the line-of-sight velocities need not stay constant over the duration of the buckling event. An increase or decrease in amplitude would alter the signal-to-noise estimate. As an order of magnitude limit, we note that the derivative of the bar strength parameter $A_{2}$ is minimized and remains constant for a duration of approximately $\sim$250 Myr (Figures 7 and 8, \citetalias{2013MNRAS.429.1949A}), which may correspond to a period during which the amplitude of the quadrupole in line-of-sight velocities is approximately constant. The finer details of timing, duration, and variation between models are left to future investigations.}
        
        \textcolor{red}{This toy model assumes that all bars evolve from being in-plane bars to being boxy/peanut bulges via the buckling instability. However, there are at least two other mechanisms for this evolution that have been proposed \citep{2014MNRAS.437.1284Q,2020MNRAS.495.3175S}, for example, a gradually increasing fraction of bar orbits getting trapped into a 2:1 frequency ($\Omega_{Z}=2 \Omega_{X}$, where $\Omega_{Z}$ is the vertical frequency and $\Omega_{X}$ is the frequency of oscillation along the bar major axis in the rotating bar frame). If these other mechanisms are common in field galaxies, for example, if they account for half of the boxy/peanut bulges, then our lower bound of 130 Myr would increase to 160 Myr.}
        
        \textcolor{red}{Secondary buckling events \citep{2006ApJ...637..214M} would also affect our estimates, as our toy model assumes that bars buckle once. \citet{2019MNRAS.485.1900S} examined the issue using a suite of 14 models. They found that secondary buckling events tend to have a longer duration and are less common. In particular, they conclude that secondary buckling is less likely to occur in galaxies with central classical bulge concentrations.}
        
      \subsection{Gas mass}
      
    Prior theoretical investigations of buckling in the literature have agreed that buckling should be suppressed in relatively gas-rich galaxies (\citealt{2006ApJ...645..209D,2007ApJ...666..189B,2009A&A...494...11W,2010ApJ...719.1470V}; \citetalias{2013MNRAS.429.1949A}). Stronger bars are preferentially found in galaxies with lower specific star formation rates \citep{2013ApJ...779..162C,2020MNRAS.491.2547R}, with observations showing that longer bars are preferentially found in galaxies with a lower specific star-formation rate \citep{fraser-mckelvie_merrifield_aragon-salamanca_masters_2019}. \cite{2012MNRAS.424.2180M} show that more bars are present in galaxies with lower HI gas fraction. Furthermore, in low-mass ($\log {M_{\rm{gal}}/M_{\odot}} \leq 10.3$) galaxies, the probability of there being a bar is a decreasing function of the central mass concentration \citep{2010ApJ...714L.260N}. 
      
    \textcolor{red}{Therefore, we expect} that the buckling instability would be most likely observed in relatively gas-poor galaxies. This expectation is perhaps weakly consistent with our results, as shown in Figure \ref{fig:gas_masses}, but the buckling galaxies appear to \textcolor{red}{follow the same general trend as the rest of the barred galaxies in the sample.} Since only two of our identified buckling galaxies have HI mass measurements, we cannot definitively conclude whether or not bar buckling is suppressed in gas-rich galaxies. The predicted trend is also absent in the results of \citet{2017MNRAS.468.2058E}. It can be seen, in their Figure 10, that the buckling galaxies have comparable gas masses as the other barred galaxies of that stellar mass. Model predictions of the effects of gas mass on buckling will only be testable once a sufficiently large number of buckling galaxy candidates are identified. 

	\section{Conclusion}
	
	\label{sec:Conclusion}
	In this paper, we present kinematic diagnostics of the bar buckling instability in both simulations and data. The main results of the theoretical work are illustrated by Figure \ref{figure:buckling_faceon} where we present our method for kinematic identification of the buckling phase. The main results of the observational work are illustrated by Figures \ref{fig:buckling_1} and \ref{fig:buckling_2}, where we present our newly identified buckling candidates. 
	
	By using N-body snapshots of a galaxy undergoing dynamical evolution, we identify a stellar velocity signal out of the plane of the disk. The signal resembles a quadrupole in shape and amplitude, and has roughly the width of the bar. We seek the buckling signal in observational data, due to its theoretically large scale: in terms of both its amplitude in stellar line-of-sight velocity, and the range in kpc over which it is detectable. To amplify the signal visually, we use unsharp masking with a circular aperture. We selected from the MANGA survey 434 galaxies which are barred and are closer to face-on than 55 degrees. We then visually identify buckling candidates based on quadrupolar disturbances of high velocity amplitude near the center of the bar.
    
    Buckling candidates are \textcolor{red}{further characterized by a} median absolute deviation fit to the simulated stellar velocity maps. We fit two free parameters -- the azimuthal angle and rotation angle in the viewing plane -- and numerous constrained parameters (inclination, bar angle, stellar velocity amplitude, and bar length). The optimization confirms that the visually identified quadrupolar signal is consistent with parameter constraints. We find five galaxies that are candidates of buckling and present examples of non-buckling galaxies for comparison. Our results suggest a lower bound for the duration of buckling to be 130 Myr, which is consistent with simulation timescales.
    
    We investigate the relationship between the HI mass fraction and stellar mass as an initial test of the hypothesis that central gas inhibits bar buckling\textcolor{red}{, but we were not able to reach a conclusion. Possible explanations for the lack of correlation may include small sample size, lack of central gas mass estimates, and inadequate treatment of gas effects by the models. Furthermore, the relationship between gas mass and stellar mass may not necessarily be fully captured by models where stellar mass is fixed}. At this time, the data are not substantive enough to form a definitive conclusion.
    
   Our results increase the number of buckling candidates discovered to date by over \textcolor{red}{twofold}, and we have developed a new approach for buckling identification reliant on both photometry and kinematics. Moving forward, there are multiple avenues for impactful follow-up: re-applying the existing methodology to MaNGA as the survey sample increases, or to other similar surveys; obtaining higher-resolution, higher signal-to-noise IFU observations of the buckling candidates; \textcolor{red}{ fitting these observations to simulated stellar velocities for more models and at more time points to better estimate the timescales involved in buckling; studying the overlap between the method developed here and that of \citet{2016ApJ...825L..30E}, which works best for moderately-inclined ($40^{\circ} \lesssim i \lesssim 70^{\circ}$) galaxies;} or applying the methodology to upcoming IFU observations of disk galaxies at higher redshift, for which the NIRSpec instrument aboard the \textit{James Webb Space Telescope} \citep{2007SPIE.6692E..0MB,2016A&A...592A.113D} will be uniquely suitable. Such future work would help enable a measurement of the buckling rate across cosmic time, and thus enable a completely new diagnostic of galaxy evolution. 

	\section*{Acknowledgments}
	
	We thank Jo Bovy and Justus Neumann for helpful feedback on the manuscript. 

    Funding for the Sloan Digital Sky Survey IV has been provided by the Alfred P. Sloan Foundation, the U.S. Department of Energy Office of Science, and the Participating Institutions. SDSS-IV acknowledges
    support and resources from the Center for High-Performance Computing at
    the University of Utah. The SDSS website is www.sdss.org.
    
    SDSS-IV is managed by the Astrophysical Research Consortium for the 
    Participating Institutions of the SDSS Collaboration including the 
    Brazilian Participation Group, the Carnegie Institution for Science, 
    Carnegie Mellon University, the Chilean Participation Group, the French Participation Group, Harvard-Smithsonian Center for Astrophysics, 
    Instituto de Astrof\'isica de Canarias, The Johns Hopkins University, Kavli Institute for the Physics and Mathematics of the Universe (IPMU) / 
    University of Tokyo, the Korean Participation Group, Lawrence Berkeley National Laboratory, 
    Leibniz Institut f\"ur Astrophysik Potsdam (AIP),  
    Max-Planck-Institut f\"ur Astronomie (MPIA Heidelberg), 
    Max-Planck-Institut f\"ur Astrophysik (MPA Garching), 
    Max-Planck-Institut f\"ur Extraterrestrische Physik (MPE), 
    National Astronomical Observatories of China, New Mexico State University, 
    New York University, University of Notre Dame, 
    Observat\'ario Nacional / MCTI, The Ohio State University, 
    Pennsylvania State University, Shanghai Astronomical Observatory, 
    United Kingdom Participation Group,
    Universidad Nacional Aut\'onoma de M\'exico, University of Arizona, 
    University of Colorado Boulder, University of Oxford, University of Portsmouth, 
    University of Utah, University of Virginia, University of Washington, University of Wisconsin, 
    Vanderbilt University, and Yale University.
    
    Calculations in this paper made use of NumPy \citep{numpy}, SciPy \citep{Scipy}, \textcolor{red}{and FastKDE \citep{OBRIEN2014222,OBRIEN2016148}}, with plotting in Matplotlib \citep{Hunter:2007} and Seaborn \citep{michael_waskom_2017_883859}.
    
    This research made use of Marvin \citep{Cherinka_2019}, a core Python package and web framework for MaNGA data, developed by Brian Cherinka, JosÃ© SÃ¡nchez-Gallego, Brett Andrews, and Joel Brownstein.
    
     K.M.X. gratefully acknowledges funding provided by Maryland Space Grant Consortium, and by Johns Hopkins University through the Dean's ASPIRE Grant. D.M.N. acknowledges support from the Allan C. And Dorothy H. Davis Fellowship, and support from NASA under award Number 80NSSC19K0589. E.A. thanks the CNES for financial support. She also acknowledges the Centre de Calcul Intensif d'Aix-Marseille and the GENCI (Grand Equipement National de Calcul Intensif) for granting her access to their high performance computing resources. 
	\bibliography{ref}

\begin{longtable*}[c]{@{\extracolsep{\fill}}llllllll@{}}
\toprule
Plate-IFU & MAD (km/s) & Fit type & $\theta$ ($^{\circ}$) & $\varphi$ ($^{\circ}$) & $\gamma$ ($^{\circ}$) & Bar len. (pix) & Vel. scale \\* \midrule
\endfirsthead
\endhead
\bottomrule
\endfoot
\endlastfoot
8947-1901$^\dagger$ & 8.65 & (buckling sim.) & 36.4 & 138.9 & 49.8 & 18 & 0.50 \\
 & 4.10 & (post-buckling sim.) & 35.8 & -40.4 & 48.8 & 20 & 0.64 \\ \midrule
8081-6103$^\dagger$ & 6.91 & (buckling sim.) & 45.8 & 20.1 & 160.0 & 14 & 0.40 \\
 & 4.97 & (post-buckling sim.) & 140.9 & -19.7 & 165.8 & 15 & 0.62 \\ \midrule
8337-1902$^\dagger$ & 8.14 & (buckling sim.) & 18.7 & -99.9 & 115.7 & 22 & 0.41 \\
 & 5.51 & (post-buckling sim.) & 160.2 & 80.0 & 130.0 & 22 & 0.50 \\ \midrule
8657-1902$^\dagger$ & 10.12 & (buckling sim.) & 48.3 & 108.6 & 40.9 & 38 & 0.63 \\
 & 8.61 & (post-buckling sim.) & 38.2 & 40.0 & -50.0 & 38 & 0.80 \\ \midrule
8451-12703$^\dagger$ & 11.26 & (buckling sim.) & 129.0 & -30.5 & 147.0 & 32 & 0.42 \\
 & 7.60 & (post-buckling sim.) & 139.6 & -40.1 & 170.2 & 34 & 0.80 \\ \midrule
8941-1901 & 7.86 & (buckling sim.) & 128.2 & -29.7 & -111.2 & 21 & 0.51 \\
 & 5.37 & (post-buckling sim.) & 59.6 & -160.2 & -119.8 & 21 & 0.65 \\ \midrule
8319-3702* & 13.99 & (buckling sim.) & 143.1 & -71.4 & -152.5 & 17 & 0.51 \\
 & 7.43 & (post-buckling sim.) & 149.2 & -50.8 & -177.3 & 19 & 0.76 \\ \midrule
8950-1902* & 9.62 & (buckling sim.) & 41.1 & -9.9 & -149.4 & 16 & 0.75 \\
 & 6.78 & (post-buckling sim.) & 135.0 & 0.0 & -128.1 & 16 & 1.00 \\ \midrule
9025-1902 & 10.78 & (buckling sim.) & 25.4 & -67.9 & 10.2 & 15 & 0.57 \\
 & 7.93 & (post-buckling sim.) & 150.8 & 39.6 & 29.2 & 16 & 0.50 \\ \midrule
8082-6102* & 11.53 & (buckling sim.) & 26.8 & -103.8 & -10.1 & 20 & 1.21 \\
 & 12.46 & (post-buckling sim.) & 25.8 & -90.0 & 10.0 & 22 & 0.97 \\ \midrule
8600-3703* & 15.18 & (buckling sim.) & 148.2 & 99.3 & 120.5 & 19 & 0.96 \\
 & 9.25 & (post-buckling sim.) & 150.0 & -89.7 & 140.1 & 21 & 1.05 \\ \midrule
8602-12705 & 9.37 & (buckling sim.) & 54.1 & -158.0 & 105.7 & 69 & 0.49 \\
 & 10.70 & (post-buckling sim.) & 132.0 & -20.0 & 110.0 & 65 & 0.90 \\ \midrule
9883-9102* & 22.48 & (buckling sim.) & 38.4 & -169.3 & 129.5 & 21 & 1.20 \\
 & 33.60 & (post-buckling sim.) & 130.0 & 0.0 & 112.8 & 21 & 1.05 \\ \midrule
9487-3704 & 12.49 & (buckling sim.) & 39.7 & 161.3 & 138.5 & 13 & 0.40 \\
 & 11.43 & (post-buckling sim.) & 141.5 & 20.1 & 133.1 & 13 & 0.45 \\ \midrule
9863-3704* & 17.18 & (buckling sim.) & 33.8 & -81.3 & -108.5 & 23 & 0.71 \\
 & 8.48 & (post-buckling sim.) & 34.2 & -38.6 & -59.9 & 22 & 0.95 \\ \midrule
8456-12701 & 6.51 & (buckling sim.) & 159.9 & 118.7 & -129.9 & 34 & 0.37 \\
 & 4.62 & (post-buckling sim.) & 20.0 & 51.9 & -139.4 & 32 & 0.40 \\ \midrule
8252-3702 & 9.25 & (buckling sim.) & 33.9 & 127.0 & 118.4 & 20 & 0.40 \\
 & 7.79 & (post-buckling sim.) & 147.7 & 30.3 & 140.3 & 21 & 0.71 \\ \midrule
8486-6101* & 14.39 & (buckling sim.) & 36.3 & 162.0 & -19.7 & 18 & 0.97 \\
 & 5.93 & (post-buckling sim.) & 123.1 & 10.9 & -10.1 & 19 & 1.00 \\ \midrule
8942-12702* & 13.72 & (buckling sim.) & 48.4 & -161.8 & 50.3 & 43 & 0.85 \\
 & 13.40 & (post-buckling sim.) & 46.0 & 9.9 & 40.2 & 40 & 1.12 \\ \midrule
9035-3704 & 7.91 & (buckling sim.) & 130.1 & -21.1 & 119.6 & 33 & 0.55 \\
 & 5.43 & (post-buckling sim.) & 136.2 & -21.1 & 115.8 & 34 & 0.92 \\ \midrule
9026-9102 & 7.72 & (buckling sim.) & 36.1 & -141.2 & -143.5 & 35 & 0.30 \\
 & 4.55 & (post-buckling sim.) & 151.7 & -59.2 & -114.8 & 39 & 0.40 \\ \midrule
8592-6102* & 12.56 & (buckling sim.) & 55.6 & -146.1 & 158.1 & 36 & 0.61 \\
 & 7.89 & (post-buckling sim.) & 139.9 & -50.0 & -170.0 & 39 & 0.86 \\ \midrule
9506-9101 & 22.11 & (buckling sim.) & 133.9 & 158.8 & 29.6 & 23 & 1.31 \\
 & 14.51 & (post-buckling sim.) & 35.7 & -160.4 & 31.5 & 22 & 1.72 \\ \midrule
9490-12702 & 13.57 & (buckling sim.) & 42.1 & 102.9 & -117.5 & 22 & 0.41 \\
 & 12.19 & (post-buckling sim.) & 141.1 & 79.7 & -120.0 & 23 & 0.40 \\ \midrule
8450-12705 & 21.43 & (buckling sim.) & 54.6 & 156.4 & -9.9 & 45 & 1.23 \\
 & 14.33 & (post-buckling sim.) & 57.9 & 163.4 & 0.0 & 45 & 1.69 \\ \midrule
9505-12704 & 8.91 & (buckling sim.) & 150.7 & -39.7 & -74.8 & 19 & 0.33 \\
 & 4.38 & (post-buckling sim.) & 153.5 & -50.8 & -54.3 & 19 & 0.37 \\ \midrule
9041-6104 & 9.19 & (buckling sim.) & 31.8 & -66.2 & -20.1 & 12 & 0.71 \\
 & 6.62 & (post-buckling sim.) & 30.6 & 129.9 & 10.5 & 12 & 0.71 \\ \midrule
8978-12702 & 25.23 & (buckling sim.) & 154.7 & 149.2 & -29.2 & 9 & 2.12 \\
 & 63.40 & (post-buckling sim.) & 27.7 & 21.7 & -30.5 & 9 & 2.17 \\ \midrule
9185-12705 & 25.28 & (buckling sim.) & 30.5 & 128.2 & -162.1 & 14 & 0.46 \\
 & 22.11 & (post-buckling sim.) & 152.2 & -140.4 & -147.9 & 14 & 0.40 \\ \midrule
8985-3702 & 9.00 & (buckling sim.) & 21.2 & -110.2 & 20.0 & 17 & 0.40 \\
 & 4.77 & (post-buckling sim.) & 162.1 & -92.3 & 59.7 & 17 & 0.44 \\ \midrule
8141-3703* & 13.17 & (buckling sim.) & 47.3 & -161.9 & 129.1 & 28 & 0.96 \\
 & 8.00 & (post-buckling sim.) & 132.1 & -10.9 & 112.5 & 27 & 1.20 \\ \midrule
8983-9101 & 17.53 & (buckling sim.) & 56.7 & -140.9 & 127.7 & 17 & 0.86 \\
 & 14.36 & (post-buckling sim.) & 36.9 & -142.5 & 133.2 & 17 & 1.31 \\ \midrule
8992-3702 & 10.27 & (buckling sim.) & 46.1 & 172.5 & -117.9 & 12 & 0.41 \\
 & 11.42 & (post-buckling sim.) & 149.9 & 30.5 & -143.7 & 12 & 0.40 \\ \midrule
8722-3702 & 8.68 & (buckling sim.) & 25.6 & -160.9 & 29.3 & 36 & 0.46 \\
 & 13.94 & (post-buckling sim.) & 25.9 & -19.6 & -179.2 & 35 & 0.76 \\ \midrule
8259-6103 & 9.47 & (buckling sim.) & 133.7 & 30.9 & 64.7 & 8 & 0.39 \\
 & 8.83 & (post-buckling sim.) & 54.5 & -30.5 & 65.5 & 8 & 0.35 \\ \midrule
9038-12703 & 7.49 & (buckling sim.) & 125.4 & -10.3 & 93.5 & 15 & 0.69 \\
 & 5.67 & (post-buckling sim.) & 120.0 & -9.8 & 89.7 & 14 & 0.73 \\ \midrule
9879-12704 & 11.93 & (buckling sim.) & 147.8 & -40.2 & -90.9 & 15 & 0.70 \\
 & 8.34 & (post-buckling sim.) & 138.4 & -38.7 & -93.7 & 14 & 0.61 \\ \midrule
8983-3704 & 22.44 & (buckling sim.) & 46.2 & -70.0 & 98.9 & 15 & 0.75 \\
 & 18.53 & (post-buckling sim.) & 141.1 & 68.6 & 82.1 & 15 & 0.70 \\ \midrule
9042-12702 & 10.00 & (buckling sim.) & 122.3 & -176.9 & -31.6 & 6 & 0.98 \\
 & 10.53 & (post-buckling sim.) & 38.8 & 172.0 & -40.4 & 6 & 1.15 \\ \midrule
8459-3703 & 9.18 & (buckling sim.) & 31.7 & 59.5 & 104.9 & 23 & 0.41 \\
 & 10.89 & (post-buckling sim.) & 144.4 & -89.5 & 143.1 & 23 & 0.41 \\ \midrule
10519-12705 & 11.02 & (buckling sim.) & 143.3 & -42.3 & 174.5 & 22 & 0.74 \\
 & 10.32 & (post-buckling sim.) & 37.5 & -129.7 & -169.3 & 25 & 0.77 \\ \midrule
8335-6101 & 20.72 & (buckling sim.) & 156.9 & -67.1 & 21.1 & 29 & 0.41 \\
 & 27.06 & (post-buckling sim.) & 30.0 & 30.0 & -19.9 & 29 & 0.40 \\* \bottomrule
\caption{Best-fit parameters for all galaxies in the sample. Galaxies deemed buckling and non-buckling are annotated with $\dagger$ and $*$, respectively. Additional parameters and initial conditions for inclination and bar position angle are shown in Table \ref{table:details}.}
\label{table:fit}
\vspace{1em}
\end{longtable*}

\begin{longtable*}[c]{@{\extracolsep{\fill}}llllllllll@{}}
\toprule
Plate-IFU & ra & dec & $b/a$ & $f_{\mathrm{DeV}}$ & $p$ & $i$ ($^{\circ}$) & bar PA ($^{\circ}$) & bar len. (px) & $q$ \\* \midrule
\endfirsthead
\endhead
\bottomrule
\endfoot
\endlastfoot
8947-1901*\^{} & 171.3836994 & 51.18040881 & 0.91 & 0.23 & 0.14 & 24.1 & 80.5 & 18 & 80.998 \\
8081-6103*\^{} & 48.0654771 & 0.2404516 & 0.72 & 0.02 & 0.12 & 44.4 & 147.6 & 14 & 42.765 \\
8337-1902*\^{} & 214.9040217 & 38.2342439 & 0.88 & 0.43 & 0.16 & 28.5 & 27.8 & 22 & 153.145 \\
8657-1902*\^{} & 9.5648540 & -0.9674056 & 0.80 & 0.88 & 0.21 & 38.2 & 96.6 & 38 & 144.681 \\
8451-12703*\^{} & 164.0289000 & 43.1565683 & 0.76 & 0.28 & 0.15 & 40.9 & 133.2 & 32 & 225.487 \\
8941-1901* & 120.7239767 & 28.4571248 & 0.66 & 0.20 & 0.14 & 49.0 & 42.1 & 21 & 58.836 \\
8319-3702 & 198.7431658 & 48.1946173 & 0.93 & 1.00 & 0.22 & 22.6 & 135.0 & 17 & 32.116 \\
8950-1902 & 194.8071141 & 27.4025448 & 0.79 & 1.00 & 0.22 & 39.1 & 46.9 & 16 & 28.994 \\
9025-1902 & 246.3581970 & 30.3094478 & 0.96 & 0.25 & 0.14 & 15.8 & 70.6 & 15$^\dagger$ & 10.242 \\
8082-6102 & 49.9459168 & 0.5845800 & 0.96 & 1.00 & 0.22 & 15.7 & 99.6 & 22 & 31.191 \\
8600-3703 & 245.8658724 & 41.7063834 & 0.95 & 1.00 & 0.22 & 17.9 & 50.7 & 20 & 7.596 \\
8602-12705 & 247.4626868 & 39.7665109 & 0.77 & 1.00 & 0.22 & 40.7 & 87.6 & 65$^\dagger$ & 70.991 \\
9883-9102 & 256.6413907 & 33.6929832 & 0.69 & 0.24 & 0.14 & 46.7 & 113.9 & 21 & 40.450 \\
9487-3704 & 123.6620484 & 44.3946431 & 0.76 & 0.13 & 0.13 & 40.7 & 148.8 & 13 & 3.604 \\
9863-3704 & 193.1711783 & 26.3658212 & 0.92 & 0.99 & 0.22 & 23.0 & 153.6 & 22$^\dagger$ & 34.474 \\
8456-12701 & 151.1912857 & 44.7057800 & 0.94 & 0.83 & 0.20 & 19.8 & 178.5 & 31$^\dagger$ & 36.256 \\
8252-3702 & 144.0598630 & 48.7456977 & 0.84 & 1.00 & 0.22 & 33.5 & 167.6 & 20 & 30.548 \\
8486-6101 & 238.0395529 & 46.3197897 & 0.73 & 0.29 & 0.15 & 43.7 & 173.0 & 19$^\dagger$ & 64.807 \\
8942-12702 & 124.0033097 & 27.0758952 & 0.83 & 0.88 & 0.21 & 34.2 & 35.0 & 41$^\dagger$ & 141.796 \\
9035-3704 & 236.6776249 & 43.3724791 & 0.82 & 0.78 & 0.20 & 35.4 & 96.9 & 33$^\dagger$ & 55.737 \\
9026-9102 & 250.8636141 & 43.3531805 & 0.86 & 0.85 & 0.21 & 31.8 & 9.7 & 39$^\dagger$ & 13.893 \\
8592-6102 & 223.3207946 & 52.0442625 & 0.74 & 0.79 & 0.20 & 43.7 & 145.5 & 39$^\dagger$ & 32.244 \\
9506-9101 & 133.5889400 & 27.1672261 & 0.82 & 0.77 & 0.20 & 35.4 & 16.5 & 22 & 362.922 \\
9490-12702 & 121.1375351 & 19.8512585 & 0.81 & 0.00 & 0.12 & 36.1 & 142.0 & 23$^\dagger$ & 21.743 \\
8450-12705 & 171.5780599 & 21.0962358 & 0.68 & 1.00 & 0.22 & 49.1 & 4.2 & 47$^\dagger$ & 42.883 \\
9505-12704 & 139.0781202 & 28.8890739 & 0.81 & 0.35 & 0.16 & 36.0 & 78.4 & 19$^\dagger$ & 35.296 \\
9041-6104 & 236.7556854 & 30.5331725 & 0.94 & 0.26 & 0.15 & 20.4 & 49.8 & 12 & 8.708 \\
8978-12702 & 249.5877689 & 40.8172086 & 0.80 & 0.05 & 0.12 & 37.0 & 132.1 & 9 & 145.814 \\
9185-12705 & 258.5824636 & 33.5488846 & 0.87 & 0.56 & 0.18 & 29.6 & 74.0 & 13 & 123.757 \\
8985-3702 & 204.0826618 & 32.8750954 & 0.86 & 0.60 & 0.18 & 31.0 & 140.6 & 16 & 92.554 \\
8141-3703 & 117.4714538 & 44.8731553 & 0.83 & 1.00 & 0.22 & 34.9 & 109.0 & 28$^\dagger$ & 30.859 \\
8983-9101 & 205.1293950 & 26.3073567 & 0.71 & 1.00 & 0.22 & 46.2 & 96.1 & 17 & 1.771 \\
8722-3702* & 125.1244429 & 51.1469357 & 0.91 & 1.00 & 0.22 & 25.3 & 17.5 & 36 & 48.433 \\
8992-3702* & 171.6572619 & 51.5730413 & 0.72 & 0.71 & 0.19 & 45.0 & 67.8 & 12 & 66.385 \\
8259-6103* & 178.8421258 & 43.7610940 & 0.76 & 1.00 & 0.22 & 42.2 & 87.4 & 8 & 32.013 \\
9038-12703* & 238.7463314 & 42.1053691 & 0.65 & 0.57 & 0.18 & 50.4 & 83.7 & 14 & 116.942 \\
9879-12704* & 198.8094968 & 26.8515209 & 0.86 & 0.60 & 0.18 & 31.3 & 57.8 & 14 & 86.356 \\
8983-3704* & 206.0079498 & 25.9411987 & 0.82 & 0.50 & 0.17 & 35.5 & 153.5 & 15 & 67.665 \\
9042-12702* & 232.0969910 & 28.2294710 & 0.66 & 0.39 & 0.16 & 49.9 & 140.6 & 6 & 30.459 \\
8459-3703* & 147.3350000 & 43.4429900 & 0.92 & 0.92 & 0.21 & 23.4 & 47.5 & 22 & 35.394 \\
10519-12705* & 155.7125565 & 4.7657940 & 0.90 & 0.87 & 0.21 & 26.5 & 137.8 & 26 & 169.851 \\
8335-6101* & 215.2292407 & 40.1210274 & 0.87 & 1.00 & 0.22 & 29.8 & 129.8 & 28 & 220.460 \\* \bottomrule
\caption{Galaxy sample with priors for inclination, bar position angle, and the size scale of the bar, in pixels. These are used for the brute force optimization and soft constraints. Last column includes lowest calculated $q$ over r=4, r=6 and r=8 unsharp filtered maps. \newline \newline \footnotesize $^\dagger$ denotes bar length measurements obtained from either of \cite{fraser-mckelvie_merrifield_aragon-salamanca_masters_2019,2011MNRAS.415.3627H}. \newline * denotes galaxies part of the original sample of 16 candidates. \newline \^{} denotes galaxies deemed buckling.}
\label{table:details}
\end{longtable*}

\label{lastpage}
	
\end{document}